\documentstyle[epsf]{article}

\newcommand{\bee}{\begin{equation}}
\newcommand{\ee}{\end{equation}}
\newcommand{\beea}{\begin{eqnarray}}
\newcommand{\eea}{\end{eqnarray}}

\newcommand{\ewxy}[2]{\setlength{\epsfxsize}{#2}\epsfbox[10 60 640 570]{#1}}

\begin{document}
\thispagestyle{empty}
\parskip=12pt
\raggedbottom

\def\mytoday#1{{ } \ifcase\month \or
 January\or February\or March\or April\or May\or June\or
 July\or August\or September\or October\or November\or December\fi
 \space \number\year}
\noindent
\hspace*{9cm} COLO-HEP-410\\
\vspace*{1cm}
\begin{center}
{\LARGE Investigating and
 Optimizing the Chiral Properties of Lattice Fermion Actions}

\vspace{0.5cm}

Thomas DeGrand,
Anna Hasenfratz,  and Tam\'as G.\ Kov\'acs\\
Physics Department, 
        University of Colorado, \\ 
        Boulder, CO 80309 USA

(the MILC collaboration)

\begin{abstract}
We study exceptional modes of both the Wilson and the clover
action in order to understand why quenched
clover spectroscopy suffers so severely from exceptional
configurations. We show that, in contrast to the case of
the Wilson action, a large clover coefficient
can make the exceptional modes extremely localized and
thus very sensitive to short distance fluctuations.
We describe a way to optimize the chiral behavior  of 
Wilson-type lattice fermion actions by studying their 
low energy real eigenmodes. We find a candidate action, 
the clover action with fat links with a tuned clover term.
We present a calculation of spectroscopy and matrix elements at
Wilson gauge coupling $\beta=5.7$. When compared to simulations
 with the standard (nonperturbatively improved) clover
 action at small lattice spacing,
the action shows good scaling behavior, with an apparent
great reduction in the number of exceptional configurations.
\end{abstract}

\end{center}
\eject

\section{Introduction}
There is an increasing accumulation of evidence from lattice simulations
of the importance of the topological
properties of the QCD vacuum \cite{VACUUM, NEGELE}.
Simulations of the pure gauge theory find that the vacuum is filled with
instantons of size 0.3-0.5 fm, with a density of about one per fm${}^{-4}$,
 although the precise numbers are still
 controversial \cite{SU3INST,SU3INST_alt}.
An elaborate phenomenology of the properties of hadrons is based on
instanton dynamics and its connection to chiral symmetry breaking \cite{IILM}.
The extent to which the complete continuum phenomenology of instanton effects
in hadronic physics is
actually realized by QCD, and can be studied using lattice simulations,
remains an open question, but its outline is present.

Lattice artifacts associated with instantons also seem to be connected to
some of the difficulties of numerical simulations with quenched QCD, 
namely, eigenmodes of the Dirac operator
which occur away from the critical quark mass, and which spoil the
calculation of the fermion propagator, 
the so-called ``exceptional configurations'' \cite{FNALINST}.
Precisely how does this occur? And, given that instantons
are responsible for chiral symmetry breaking, is it possible to optimize
the lattice discretization of a fermion action with respect to its topological
properties? Our   goal is to shrink the range of bare quark mass
over which  the low lying real eigenmodes of 
the Dirac operator occur,  on background gauge field configurations
which are typical equilibrium configurations
at some gauge coupling.

A chiral improvement of the lattice Dirac operator is
expected to
decrease the additive mass renormalization
 of zero-modes as well as their spread. This
is indeed the case for classically perfect actions which
have been proven to have exact zero modes with no shift or
spread at all \cite{HNL}. On the other hand, surprisingly, this
does not seem to be true for the simplest improvement on the
Wilson action, the clover action. Although the clover term
reduces the additive mass renormalization as compared to the
Wilson action, the problem with exceptional configurations
appears to be more severe for the clover than for the Wilson
action; at least with the non-perturbatively determined value
of the clover coefficient $C_{sw}$ \cite{DESY96173}. On a set of
50 12$^3\times$24 $\beta=5.9$ configurations the Fermilab
group found 7 exceptional modes both with the Wilson and the
clover action $(C_{sw}=1.91)$ but the spread of the clover modes
in terms of the pion mass
was at least twice that of the Wilson modes. This
indicates that surprisingly, improving the action makes
the situation worse for exceptional configurations.
In the present paper we show why this happens by studying
how the real modes of the Dirac operator change with the clover
coefficient. 

We are then drawn to an alternate action,
a modified clover action, where
the gauge connections are replaced by APE-blocked \cite{APEBlock}
 links, and the
clover coefficient $C_{sw}$ is tuned to optimize chiral properties.
Specifically,
\beea
S & = &\sum_n (m+4)\bar \psi(n) \psi(n) \nonumber  \\
& - & {1\over 2}  \sum_{n \mu}(\bar \psi(n)
(1 - \gamma_\mu) V_\mu(n) \psi(n+ \mu) 
+
\bar \psi(n)(1 + \gamma_\mu) V_\mu^\dagger(n-\mu) \psi(n - \mu)
\nonumber  \\
& + & {C_{sw} \over 2} \sum_{n,\mu,\nu} \bar \psi(n) i \sigma_{\mu\nu} 
P_{\mu\nu}(n) \psi(n)
\label{FATCLOVER}
\eea
with  
\beea
V^{(n)}_\mu(x) = &
(1-\alpha)V^{(n-1)}_\mu(x) \nonumber  \\
& +  { \alpha\over 6} \sum_{\nu \ne \mu}
(V^{(n-1)}_\nu(x)V^{(n-1)}_\mu(x+\hat \nu)V^{(n-1)}_\nu(x+\hat \mu)^\dagger
\nonumber  \\
& +  V^{(n-1)}_\nu(x- \hat \nu)^\dagger
 V^{(n-1)}_\mu(x- \hat \nu)V^{(n-1)}_\nu(x - \hat \nu +\hat \mu) ),
\label{APE}
\eea
with  $V^{(n)}_\mu(x)$  projected back onto $SU(3)$ after each step, and
 $V^{(0)}_\mu(n)=U_\mu(n)$ the original link variable. 
We take $\alpha=0.45$ and $N=10$ smearing steps,
chosen because of our previous work on instantons \cite{SU3INST}.
This choice of parameters is not unique and might not even be optimal.
$P_{\mu\nu}$ is the usual clover set of links,
but built of the $V_\mu$'s. At Wilson gauge coupling $5.7-5.8$, the
best choice is $C_{sw}=1.2$, and it decreases to the tree-level $C_{sw}=1$ value
at larger $\beta$.
We will also quote simulation data in terms of
 the hopping parameter, $\kappa=1/(8+2m)$.

The optimized  action has a spread 
of  low lying real eigenmodes with respect to the
bare quark mass which is less than a third  of the spread 
for the usual Wilson action.
In terms of the square of the pion mass the spread is about three times
smaller for the optimized action than
for the standard Wilson action.
The action has other good features, as well: 
the renormalization factors connecting
lattice quantities to their continuum values appear to be very close
to unity.  The action also appears to require only about half the
number of sparse matrix inversion steps as the usual clover action 
(at equivalent values of the physical parameters).

The action we propose is completely unimproved in its kinetic properties,
so its dispersion relation and heavy quark mass artifacts are identical to
those of the Wilson action.  It would be very easy to improve it by
beginning with a more complicated free fermion discretization.

Fermion actions with fat links have received considerable attention
in the past year.
The first use of them we know of (although  with
a  different motivation) was in the simulations of QCD on
cooled gauge fields by the MIT group \cite{MITCOOL}.
The MILC collaboration \cite {MILC}, Orginos and Toussaint \cite{ORT}
and 
Laga\"e and D.~K. Sinclair\cite{SINCLAIR} have shown that modest fattening
considerably improves flavor symmetry 
restoration in simulations with staggered fermions.
We have used calculations of staggered spectroscopy on highly smoothed
gauge configurations
to compare chiral symmetry breaking  in 
$SU(2)$ gauge theory and in instanton backgrounds \cite{COLOINST}.
Finally, all fixed point actions and approximate fixed point  actions
we know of \cite{ALLFP,HYPER}
for fermions seem to incorporate fat links.
Fixed point fermions realize the index theorem and retain
chiral symmetry at nonzero lattice spacing \cite{HNL}, and so one
way of viewing a fat link action is as an approximate FP action,  which
includes its chiral properties but not its kinetic ones.

An apparent drawback of fat link actions is the lack of a transfer matrix
between consecutive time slices. N APE steps can mix links up to $\pm N$
timeslices away and a strict transfer matrix cannot be defined on time slices
closer than $2N$ lattice spacings (20 in our case).
In practice we are concerned only with the exponential decay of correlation
functions and we found asymptotic decay after 2-5 time slices in our
simulation. This is expected if we realize that APE smearing is basically a
random walk whose range can be estimated as $\sqrt{N} \alpha\approx 1.4$
in our case.

The outline of the paper is as follows: In Section 2 we review the
continuum index theorem and describe how lattice fermions fail to
reproduce it. We then describe how we find the real eigenmodes
of the Wilson-Dirac operator.
All this is basically a review, and experts may skip it.
In Section 3 we describe the connection between small topological
objects and exceptional configurations for the thin-link clover
action.
In Section 4 we return to the fat link clover action and
describe the tests we performed to tune the action.
Section 5 describes a calculation of spectroscopy and matrix elements 
at $aT_c=1/4$ (Wilson gauge coupling $\beta=5.7$) using the new action.

\section{Measuring the Chiral Properties of Wilson-like Fermions}

The reader might recall that in the continuum the local topological density
\bee
q(x) = {1\over {32\pi^2}}\epsilon_{\mu\nu\rho\sigma}F_{\mu\nu}F_{\rho\sigma}
\ee
is related to the divergence of the flavor singlet (with $n_f$ flavors)
 axial-vector current  
\bee
\partial_\mu \bar \psi i \gamma_\mu \gamma_5 \psi =
 2 m \bar \psi i \gamma_5 \psi + 2in_f q.
\ee
Integrating over all $x$, this relation implies
 a connection between the topological charge $Q=\int d^4x q(x)$
and a mode sum,
\bee
Q = m {\rm Tr} \bar \psi \gamma_5 \psi =
 m {\rm Tr} \gamma_5 {1\over{ \gamma \cdot D + m}} = m \sum_s 
{{f^\dagger_s \gamma_5 f_s}\over{i \lambda_s + m}}
\ee
where $f_s$ are the eigenfunctions of the (antihermitian) Dirac operator
$\gamma \cdot D$, $\gamma \cdot D f_s = i 
\lambda_s f_s$, with $\lambda_s$ real.
The property $\{\gamma_5 , \gamma \cdot D\}=0$ leads to the condition that
$f_s^\dagger \gamma_5 f_s =0$ if $\lambda_s \ne 0$, and if $\lambda_s=0$
we may choose $f_s$ to have a definite chirality, $\gamma_5 f_s = \pm f_s$.
Thus it follows that
\bee
Q = \sum_{s, \lambda_s=0} = f_s^\dagger \gamma_5 f_s = n_+ - n_-
\ee
where $n_+$ and $n_-$ are the number of 
zero eigenmodes with positive and negative chirality. 
This is the index theorem.

On the lattice, essentially every statement in the preceding paragraph
 is contaminated by lattice artifacts \cite{SMITVINK}. 
Here we focus on Wilson-like
fermions, where chiral symmetry is broken by the addition of terms
proportional to the Dirac scalar and/or tensor operators.
The Wilson  or clover fermion
action analog of $\gamma \cdot D$, $D_w$, is neither
Hermitian nor antihermitian and its eigenvalues are generally complex.
$D_w$ can also have real eigenvalues. These real 
eigenvalues usually do not occur at zero. 
Their locations spread around $-m_c$ where $m_c$ is the critical 
bare mass, at which the pion mass vanishes. For Wilson-like
fermions $m_c$ is usually negative and its magnitude decreases
with increasing clover term.

On smooth, isolated instanton background configurations,
the location of the real eigenmode varies with the 
size $\rho$ of the instanton. For large instantons, 
the low lying real mode occurs close to zero.
Accompanying this near-zero mode, there is a set of modes which 
do not cluster around zero, but around
$\simeq O(1)$ (in lattice units).  These are ``doubler modes.''  
As one decreases the instanton size, the eigenmode near zero 
shifts in the positive direction, approaching the doubler modes. 
This $\rho$-dependent mass shift and the 
location of the doubler modes all depend on the particular choice of lattice
fermion action \cite{SIMMA,SCRI_smooth}.
The shift of the low lying real eigenmode for a given instanton size for
the Wilson action is larger than for the clover
action, which has better chiral properties. Nevertheless, for both actions,
as the instanton size decreases,  sooner or later the lowest eigenvalue
starts to increase, approaches the doublers and 
eventually annihilates with one of the doublers - the fermion 
does not see the instanton any longer.

\begin{figure}[h!tb]
\begin{center}
\vskip 10mm
\leavevmode
\epsfxsize=100mm
\epsfbox{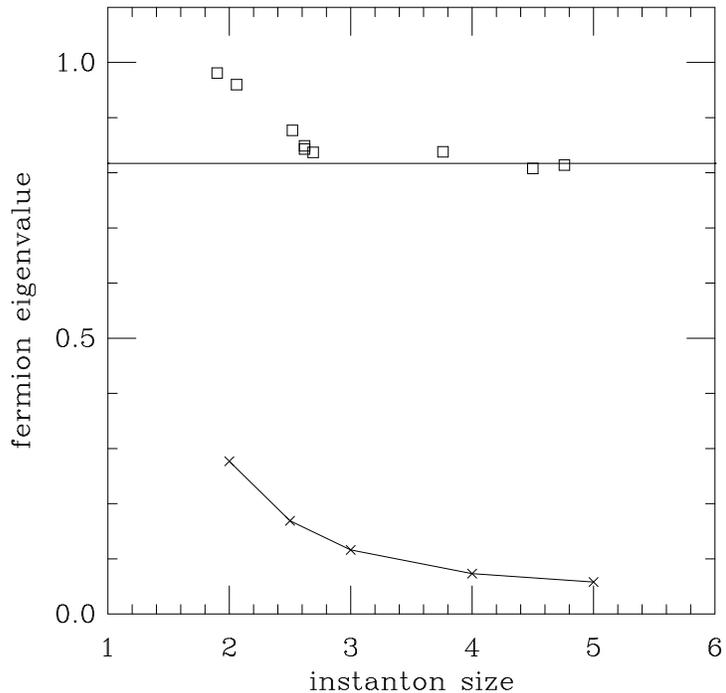}
\vskip 10mm
\end{center}
\caption{The instanton size (in units of the lattice spacing)
vs.\ the eigenvalue of the
corresponding  (Wilson action) fermionic real mode on smooth instantons
(crosses) and on real Monte Carlo generated configurations
at $\beta=6.0$ (squares). The horizontal line indicates
$-m_c$ for the $\beta=6.0$ quenched ensemble.}
\label{fig:rho_vs_pole}
\end{figure}

Real eigenmodes of the Dirac operator also occur on equilibrium
 background configurations. Typically, they occur around some average value,
an overall $\beta-$dependent mass shift. The real eigenmodes also sense the
topological content of the background configuration and their positions
can sometimes be correlated with the sizes of the background instantons.
This  is illustrated in Fig.\
\ref{fig:rho_vs_pole} for the Wilson action, where we show the instanton
size versus the corresponding fermionic eigenvalue both
for smooth instantons and instantons identified on real Monte Carlo
generated configurations at $\beta=6.0$ on $12^4$ lattices.
The instanton sizes were measured using the method of Ref.\
\cite{SU3INST} and the real modes were extracted as
described below. To find the corresponding instanton
for each real mode, we compared the quark density of the given
mode to the profile of the charge density. Usually the quark
density had a well identifiable lump sitting exactly on an
(anti)instanton.
These lattices are fairly small in physical units with volumes
$V \simeq 2$ fm${}^4$ and contain on average about two topological
objects.
In some cases, however,
these modes were peaked at not only one instanton, but more,
and sometimes the quark density was so spread out that it was
hard to identify a well defined peak.
Fig.\ \ref{fig:rho_vs_pole} contains results
only from uniquely identifiable modes.

In addition to the low lying modes there are also doublers.
Again there are two issues: First, do the
low lying modes separate from the doubler modes, and second, what
is the spread of the low lying modes.

The first question is important because, if the low lying modes do
not separate from the doubler modes, then the physics of chiral symmetry
breaking with a lattice cutoff is different than in the continuum.
The lattice theory is no longer an  approximation to the
continuum,
and one cannot speak of chiral symmetry breaking as being induced by
instantons.  Chiral symmetry is certainly broken in QCD for
any value of the cutoff, including the strong coupling limit
\cite{STRONG},
but the mechanism does not involve instantons.

Note that if the low lying modes do not separate from the
doubler modes, it does not make sense to talk about the spread of
the low lying modes either.

The determination of the real eigenvalues $\lambda$ of $D_w$ is equivalent 
to computing the bare mass values $m_\lambda= -\lambda$, where
the quark propagator $(D_w+m)$ becomes singular. In the remainder of
the present paper instead of the eigenvalue, we shall always 
quote the corresponding bare mass value, where the propagator 
is singular. On any particular  background gauge configuration this
can be computed very simply.
We approximate
\bee
P(m) = \langle \bar \psi \gamma_5 \psi\rangle
 = {\rm Tr} (D_w+m)^{-1}\gamma_5
\ee
with a noisy estimator: cast a random vector $\eta_i$ on each site
$i$, solve $(D_w+m) \chi = \eta$ for $\chi$ and measure
\bee
P(m)_\eta = \sum_{ij} {\rm Tr} \bar \eta_i \gamma_5 \chi_i
\label{PG5P}
\ee
as a function of $m$.
If an eigenmode of $D_w$ is located at $\lambda$, its presence is signaled
by the appearance of a pole in $P(m)$ at $m=-\lambda$.
This is a variant on the standard method of measuring $\langle \bar \psi \psi
\rangle$ in a lattice simulation. It has also 
seen considerable use by the Fermilab group \cite{FNALINST}
in their studies of exceptional configurations (they use a flat source
$\eta_i=$ constant, not a noisy source).
The eigenfunction itself of a particular mode can be 
found by performing the inversion for a test mass very close
to the given pole. Then $(D+m)^{-1}$ projects out the corresponding
eigenmode from the source. In this case $\chi^\dagger \chi$ is
(proportional to) the probability density for that mode. For computing
the probability density we used a flat source rather than a random
one.

\section{Trouble with the Thin Link Clover Action}

\begin{figure}[h!tb]
\begin{center}
\vskip 10mm
\leavevmode
\epsfxsize=100mm
\epsfbox{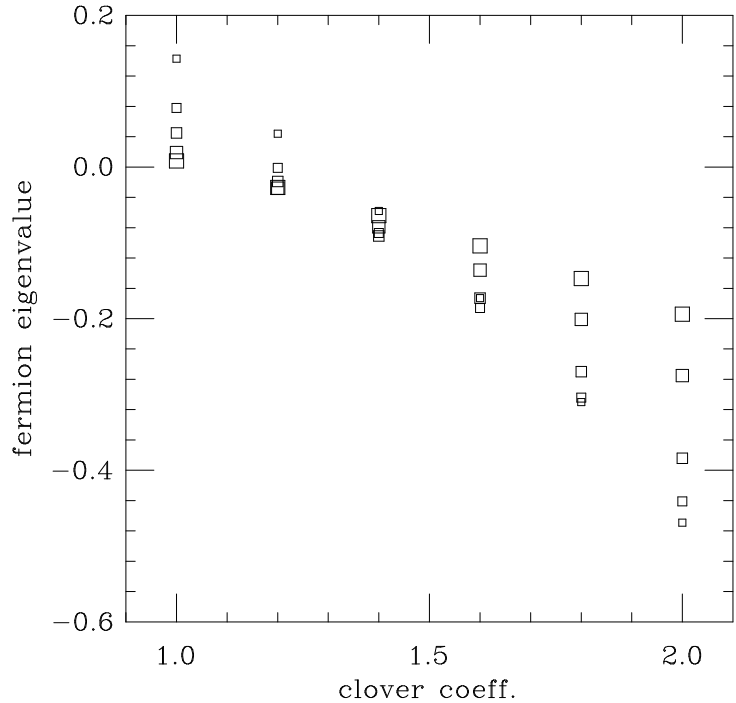}
\vskip 10mm
\end{center}
\caption{The real fermionic eigenvalue of
the clover action versus the clover coefficient
on instantons of sizes $\rho/a=1.2, 1.4, 1.6, 2.0, 2.5$. Bigger symbols
correspond to larger instantons.}
\label{fig:clover}
\end{figure}

As we have seen in Fig.\ \ref{fig:rho_vs_pole}, 
on fine enough lattices with the Wilson
action there is a strong correlation between the
instanton size and the eigenvalue of the corresponding fermionic
real mode; (near) exceptional modes correspond to large
instantons. We shall now explore what happens in the presence
of the clover term. We start by looking at smooth single instanton
configurations. Our naive expectation is that since the
clover term is designed to improve the chiral properties of the
action, its inclusion
should make the fermion modes more ``continuum-like''.
In other words, it should move the modes corresponding
to different instanton sizes closer to one another and also
shift them all towards zero. This is indeed what happens.
In Fig.\ \ref{fig:clover} we plot how the fermion eigenvalues
associated with instantons of various sizes
change as a function of the clover coefficient. For better
legibility in the figure we include only the range of
$C_{sw} \geq 1.0$. For the Wilson action $(C_{sw}=0)$ the
eigenvalues corresponding to instantons of size $1.2 \leq \rho
\leq 2.5$ are spread between 0.19 and 0.86.

   As $C_{sw}$ increases from zero to the tree level value, $C_{sw}=1$,
all the modes move closer
to zero and their spread decreases. If $C_{sw}$ is further increased,
the modes start to pass through zero and also the trajectories
corresponding to instantons of different sizes start to cross
one another. By $C_{sw}=1.4$ the order of all the eigenvalues
for instanton sizes $\geq 1.5a$ has been reversed. As $C_{sw}$
is increased, modes of smaller and smaller instantons cross over
to the other side of the distribution.

What happens on equilibrium Monte Carlo configurations?
We first performed a detailed study of the
Fermilab quenched $\beta=5.9$ exceptional
configurations reported in Ref.\ \cite{FNALINST}.
We determined the wave function of both the seven Wilson
and the seven clover exceptional modes together with the instanton
content of the corresponding background
gauge configurations. Whereas in most of
the cases the Wilson modes were sitting on a single large instanton,
the clover modes were always spread over several topological objects.

This suggests that the mechanism responsible for the exceptional
configurations might be different for the Wilson and the clover action.
If indeed there is such a difference, we expect it to be more
pronounced on coarser lattices, where exceptional configurations
are more of a problem and also the clover coefficient is usually set
to larger values. To explore this further, we studied an ensemble of
$\beta=5.7$ $6^3\times16$ lattices.

We followed how the wave function of a few typical very exceptional
modes changed with the clover coefficient starting from the
Wilson action $(C_{sw}=0.0)$ up to the non-perturbatively determined
value $(C_{sw}=2.25)$. Ideally, we would have liked to compare
the peaks of the quark density in the wave function with the
instanton locations and sizes. Unfortunately, at this rather large
value of the lattice spacing the instantons are quite small and the
fluctuations large, so the size of the instantons is not well defined.
We can still ask however how the fermions with the Wilson action
``feel'' whatever is left of the topology on these coarse lattices.
We have therefore computed the wave function of the real modes
in the physical branch with the Wilson action. For each wave
function we also determined the position of the largest
peak and its width. Usually the magnitude of the
 second largest peak was much smaller.
The precise definition of the width is irrelevant here. Since we
use the width only to qualitatively compare the localization
of different wave functions, any sensible definition would give
essentially the same results. We summarize the data for
one typical configuration in Table \ref{tab:Wilson_wfn}.

\begin{table}
\begin{tabular}{|c|l|l|l|l|l|}
\hline
   eigenvalue & location  & width $(a)$
                      & $\psi^{\dagger} \psi$ max/min \\
\hline
   0.999 &    2 1 0 6  & 2.8 & 1.8$\times$10$^2$  \\
   1.041 &    4 2 1 3  & 3.3 & 1.4$\times$10$^2$  \\
   1.241 &    0 5 5 0  & 2.3 & 5.3$\times$10$^3$  \\
   1.530 &    2 1 2 14 & 2.1 & 2.1$\times$10$^5$  \\
   1.610 &    0 0 5 1  & 1.4 & 5.8$\times$10$^5$  \\
\hline
\end{tabular}
\caption{The physical real eigenmodes of the Wilson Dirac
operator ($C_{sw}=0.0$) on a typical $\beta=5.7$ $6^3\times 16$ lattice.
We also show the location where the quark density is peaked,
the width of the peak (in lattice units)
and the ratio of the maximum and minimum
quark density in the given mode over the whole lattice.}
\label{tab:Wilson_wfn}
\end{table}

The critical mass is $m_c=-1.04$, so the first mode is exceptional.
The last mode in the Table is approximately halfway towards the
doublers. There is a clear tendency that the quark density in
modes close to $-m_c$ has an extended broad peak and for modes
towards the doublers the peaks get narrower and sharper. This is
in complete agreement with \cite{SCRI_spectralflow}.
Let us now follow the continuous evolution of the Wilson
mode at 0.999 if the clover term is gradually turned on.
Up to $C_{sw}=1.0$ the main peak of the quark density remains
at the same location. It only slowly gets narrower: at
$C_{sw}=1.0$ its width is 2.3. Increasing $C_{sw}$ further,
another peak appears and its relative significance
increases with the clover coefficient. By $C_{sw}=2.25$
the wave function is completely concentrated on a very
sharp peak (of width $\approx 1$) located at (0 0 5 1).
This is the location of the $\lambda=1.610$ Wilson-mode
which is halfway between the doublers and $-m_c$.
The wave function of this $C_{sw}=2.25$ exceptional
mode is very similar to that of a Wilson mode lying halfway between
the physical and the doubler branch.

We can now easily put the picture together and describe
qualitatively how the exceptional modes change with the clover
coefficient.  The first Wilson mode
can be roughly thought of as being concentrated on a single large
 instanton (or instantons with roughly equal size).
As the clover term is turned on,
the zero modes corresponding to different instanton sizes get closer and the
quark wave function spreads
over several topological objects, as we also saw on the Fermilab
configurations. If $C_{sw}$ is further increased,
the zero modes separate again,but this time the well localized
modes (corresponding to smaller instantons) have smaller eigenvalues. 
The relative
importance of the broader peaks decreases and the mode can become
entirely concentrated on a very sharp peak. We want to emphasize
however that on a given ensemble of gauge configurations at a given
value of $C_{sw}$ different (near) exceptional modes can look
qualitatively very different. Some of the modes are entirely
concentrated on a sharp peak --- typically these are the most exceptional
ones --- some have several peaks of various widths. The exact
picture presumably depends on the arrangement of topological
objects and fluctuations on the given configuration.

The sharply peaked modes look very much like (near) doubler
modes of the Wilson action and
the corresponding eigenvalues and the way they change
with the clover term can be very sensitive to the fluctuations
on the shortest distance scale.
We indeed encountered several exceptional eigenvalues at
$C_{sw}=2.25$ (the non-perturbatively determined value)
 that moved anomalously quickly with the
clover coefficient. This is the reason why the clover
coefficient cannot be optimized to minimize the spread of the
real eigenvalues.

\section{Testing and Tuning Actions}

 Our goal is still to tune a clover action to have good chiral
properties, i.e. to minimize the spread of the
low lying modes. We see that if the clover coefficient is made too
large,
that program will fail. 
We also know from previous 
work \cite{COLOINST} that an action with fat
links is insensitive to  short distance fluctuations, 
but still knows about instantons and the additional long
distance behavior of the gauge field responsible for confinement. 
So we will begin with a fat link action of Eqn. \ref{FATCLOVER}.
Throughout this paper we create the fat link by 10 APE smearing 
steps with smearing coefficient $\alpha=0.45$.

\begin{figure}
\centerline{\ewxy{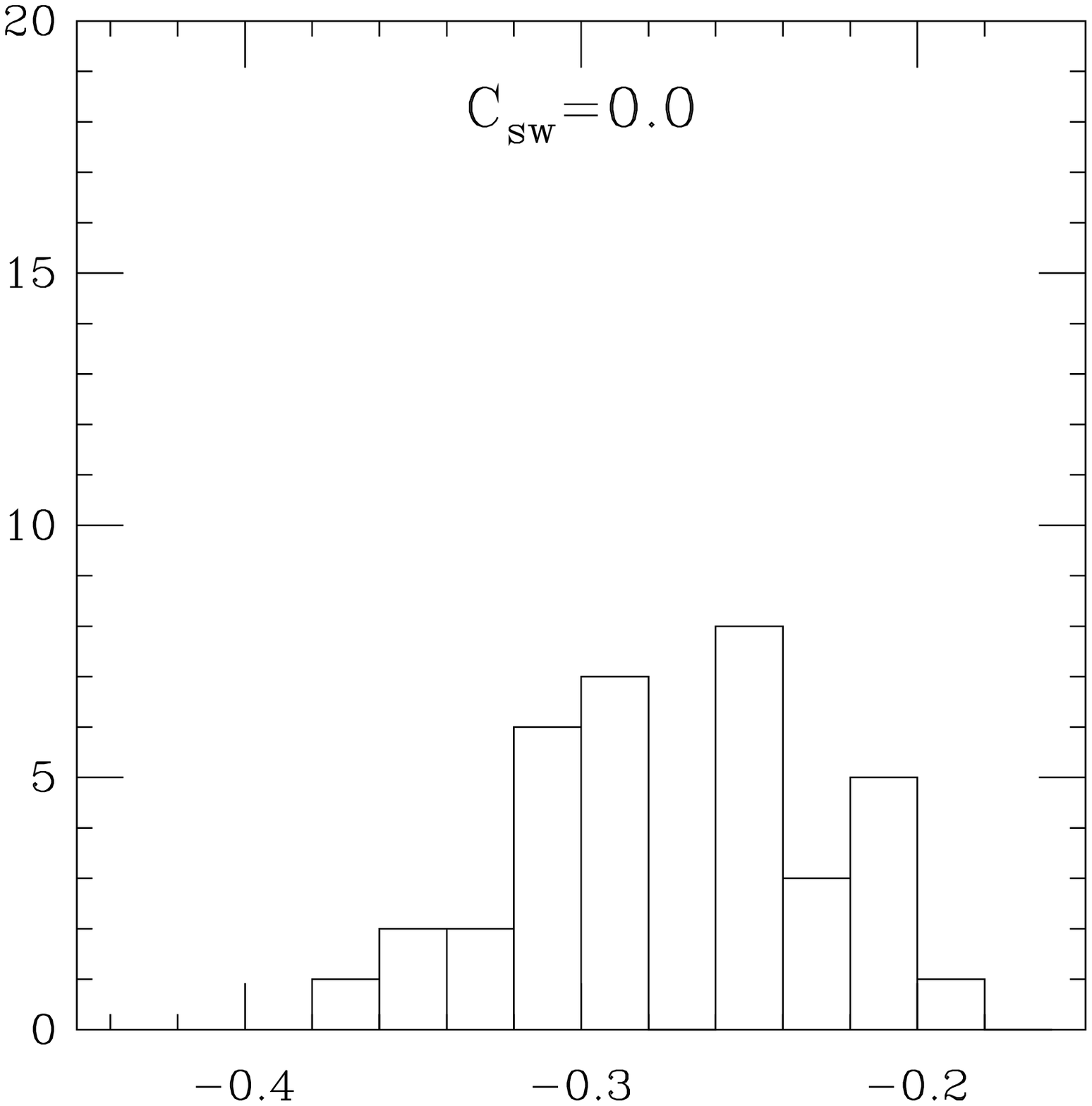}{80mm}
\ewxy{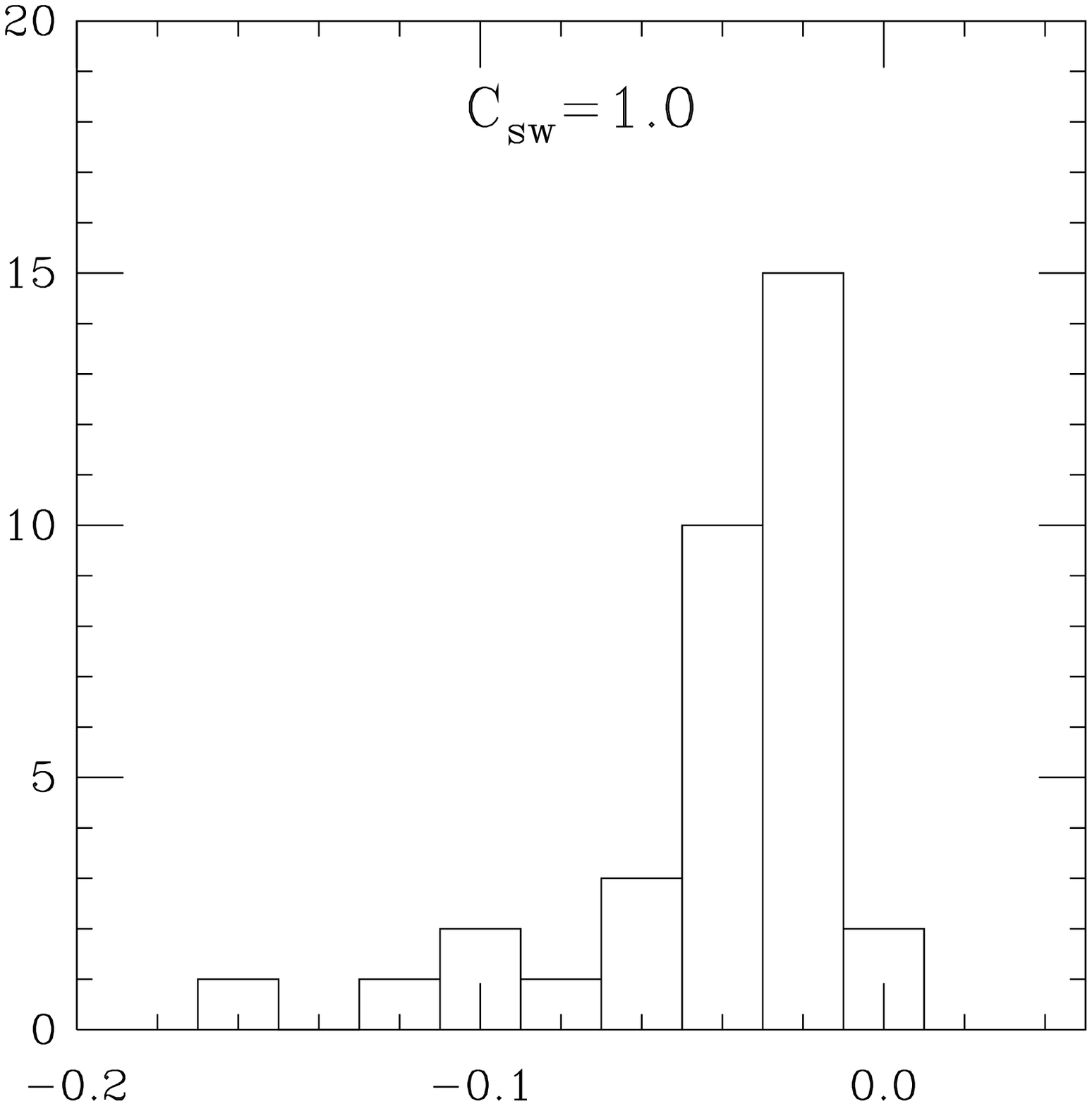}{80mm}}
\vspace{0.5cm}
\centerline{\ewxy{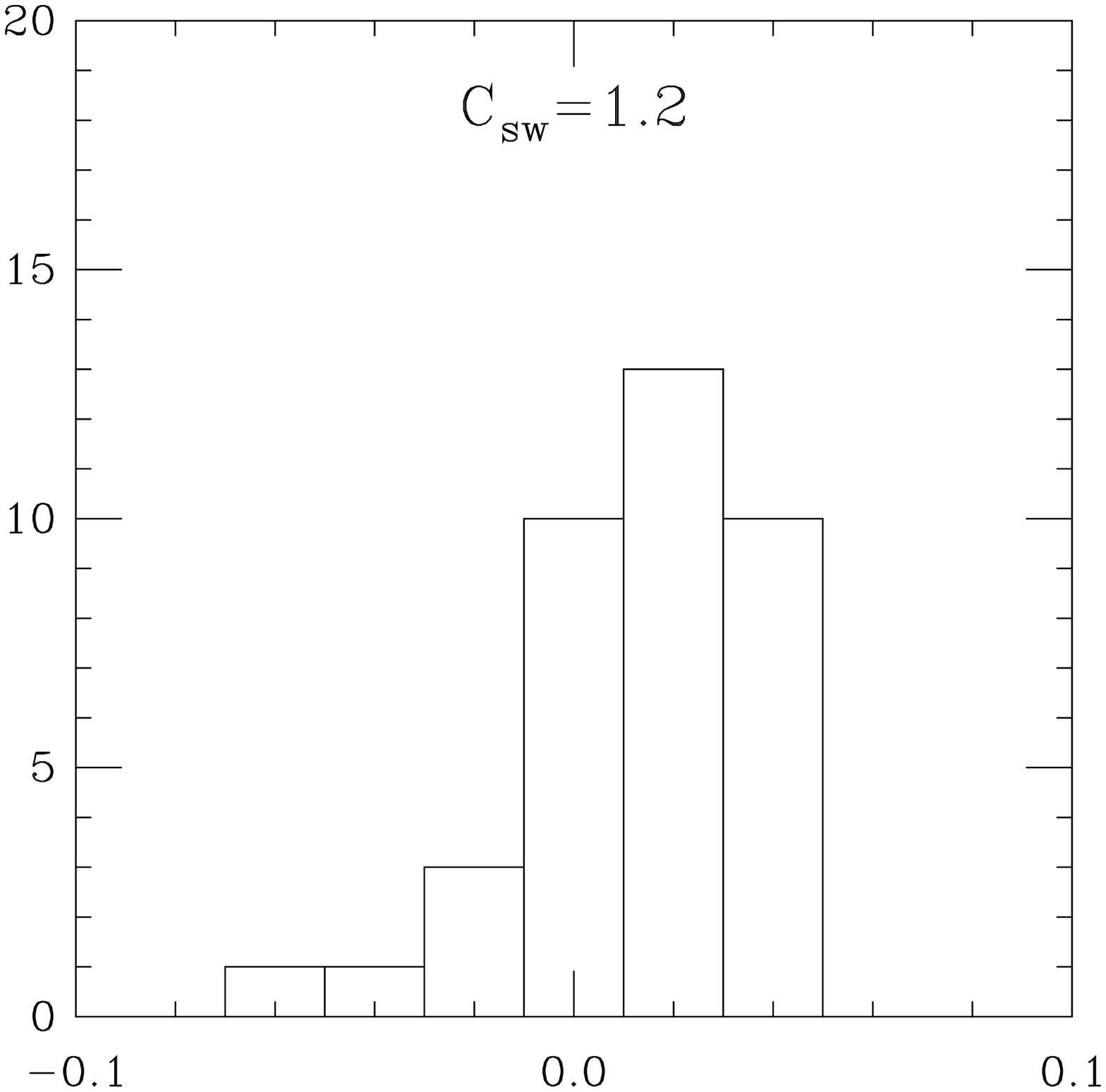}{80mm}
\ewxy{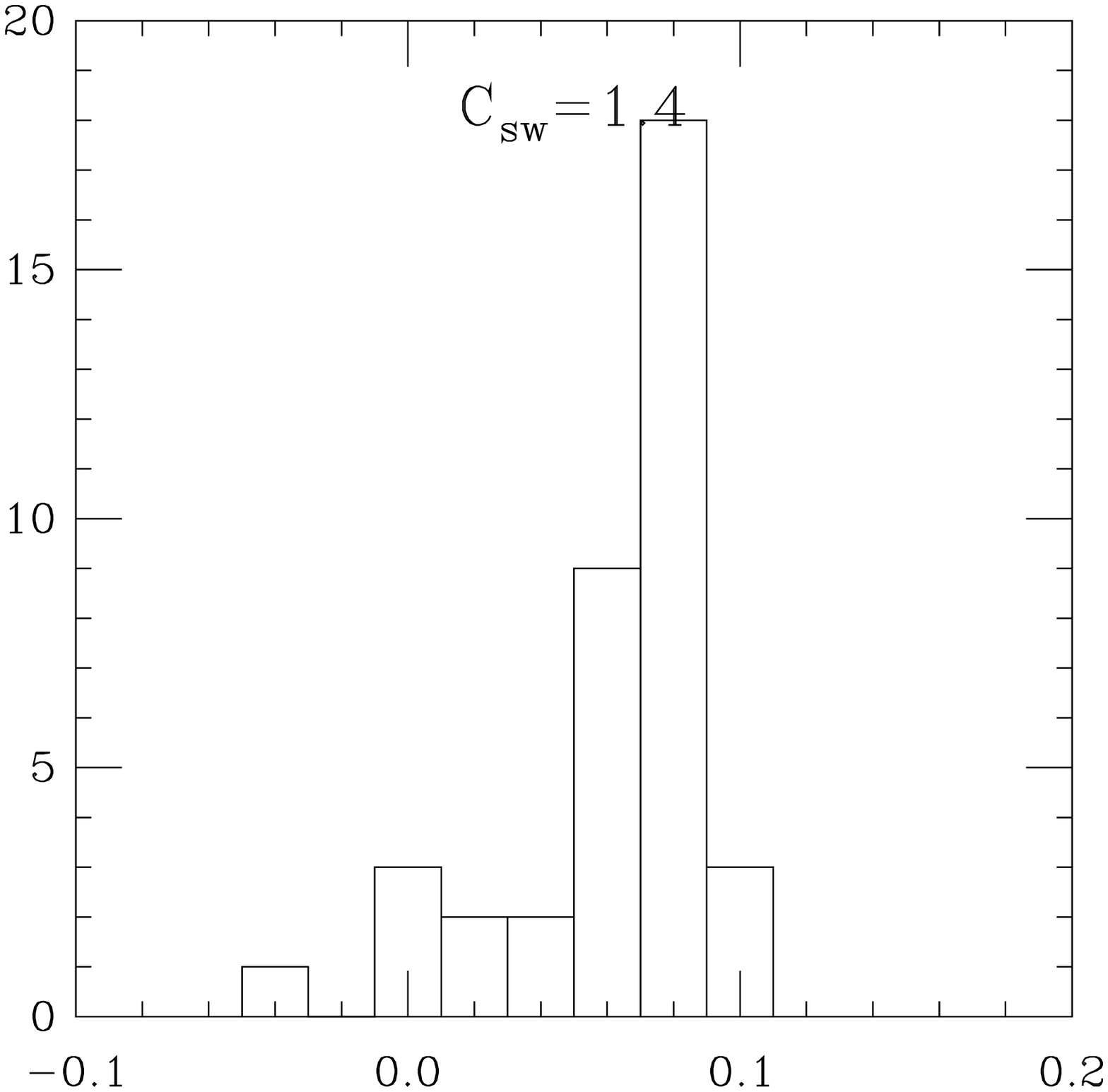}{80mm}}

\caption{The distribution of the bare masses $m$, where the 
quark propagator is singular at $\beta=5.8$  
of fat link Wilson (a); fat link clover
fermions $C_{sw}=1.0$ (b);  $C_{sw}=1.2$ (c); $C_{sw}=1.4$ (d).}
\label{fig:polevsclover5.8}
\end{figure}

Fig. \ref{fig:polevsclover5.8}  shows the distribution of the real 
eigenmodes at $\beta=5.8$ in terms of the bare quark mass, where 
the propagator is singular, for clover coefficients $C_{sw}=0.0$ 
(Wilson action), $C_{sw}=1.0$, $C_{sw}=1.2$ and $C_{sw}=1.4$.
 The poles
were located using the pseudoscalar density function of 
Eqn.\ \ref{PG5P} on 40 $8^4$ configurations.
The horizontal ranges of the histograms  are 
equal to the range the eigenmode search had been performed. 
To the right of the distributions is the  confining phase. 
If one  measures the pion mass in the positive
mass region and extrapolates $m_\pi^2$ to zero with the bare mass $m_0$, 
one finds that the
bare mass at which $m_\pi=0$ lies at a value $m_c$ located within
the range of  the low-lying real eigenmodes.
 If a particular configuration
has an eigenmode with $m_\lambda > m_c$,  the quark 
propagator will be singular here.
To the left of the distributions, in the negative quark 
mass region, are the doublers. In Fig. 
\ref{fig:polevsclover5.8} there is no indication of them, 
since the doublers did not show up within the range
we scanned for eigenmodes. It appears that the doublers 
and low lying modes are well separated at
$\beta=5.8$.

Using several fermion actions --- all different from the one in the present 
paper ---  the authors of Ref.\ \cite{SCRI_spectralflow}
found that real modes cover the whole investigated negative quark mass
 range, for all values of gauge couplings they studied, as long as the volume
is large enough.
Our conclusions are not in contradiction with this result.
They are much more modest--that the action can be tuned to make the
distribution of real eigenmodes sharply peaked near a small mean value,
and that the doublers are well separated.
No intrinsic property of our action protects chiral symmetry at nonzero
lattice spacing and so in principle it should have real eigenmodes anywhere
 as well.
But for practical calculations there is a big difference between
a distribution with narrow peaks or a uniform spread of eigenmodes, and
 we are trying to construct the first kind of action.

\begin{figure}
\centerline{\ewxy{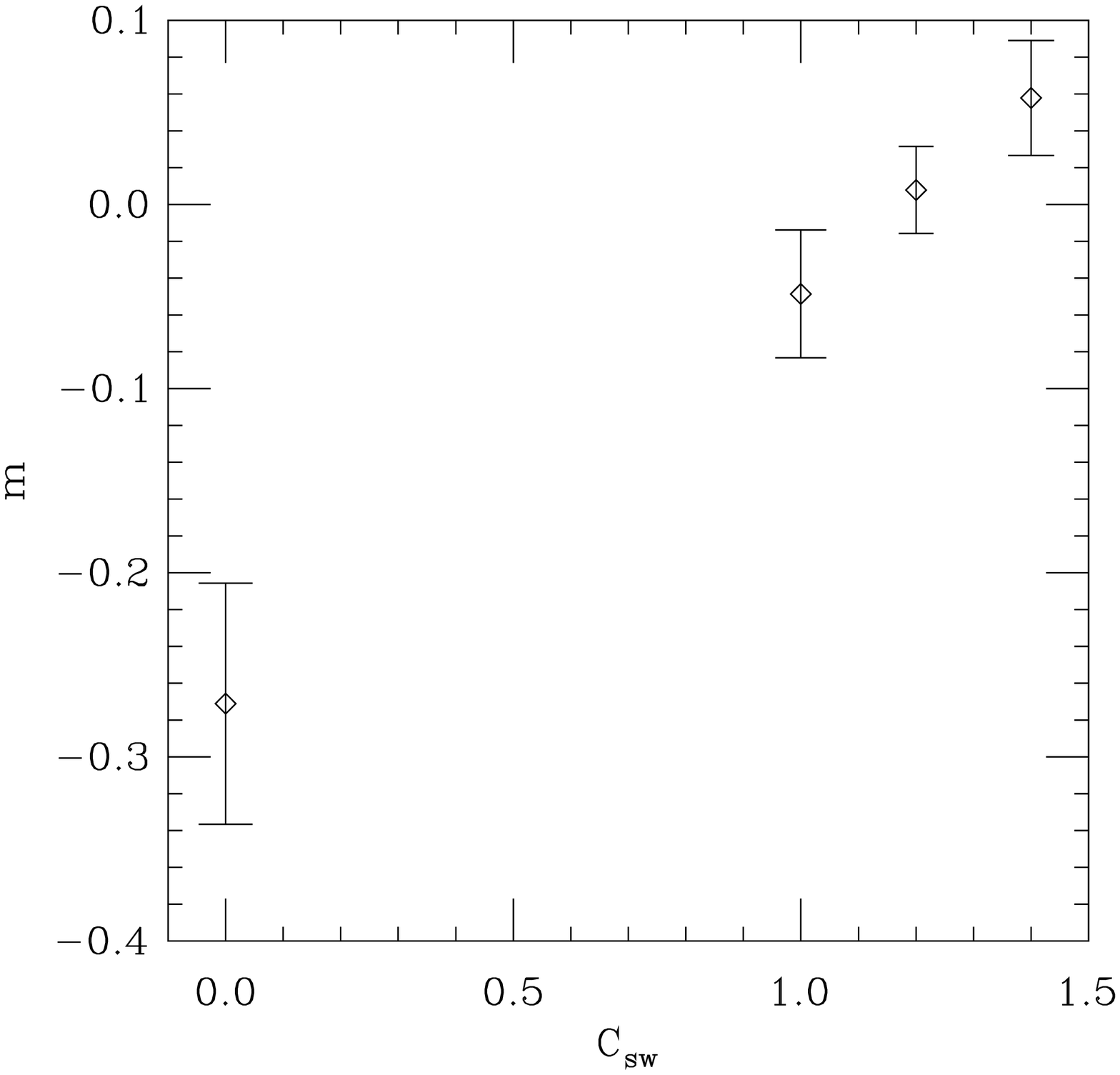}{80mm}
\ewxy{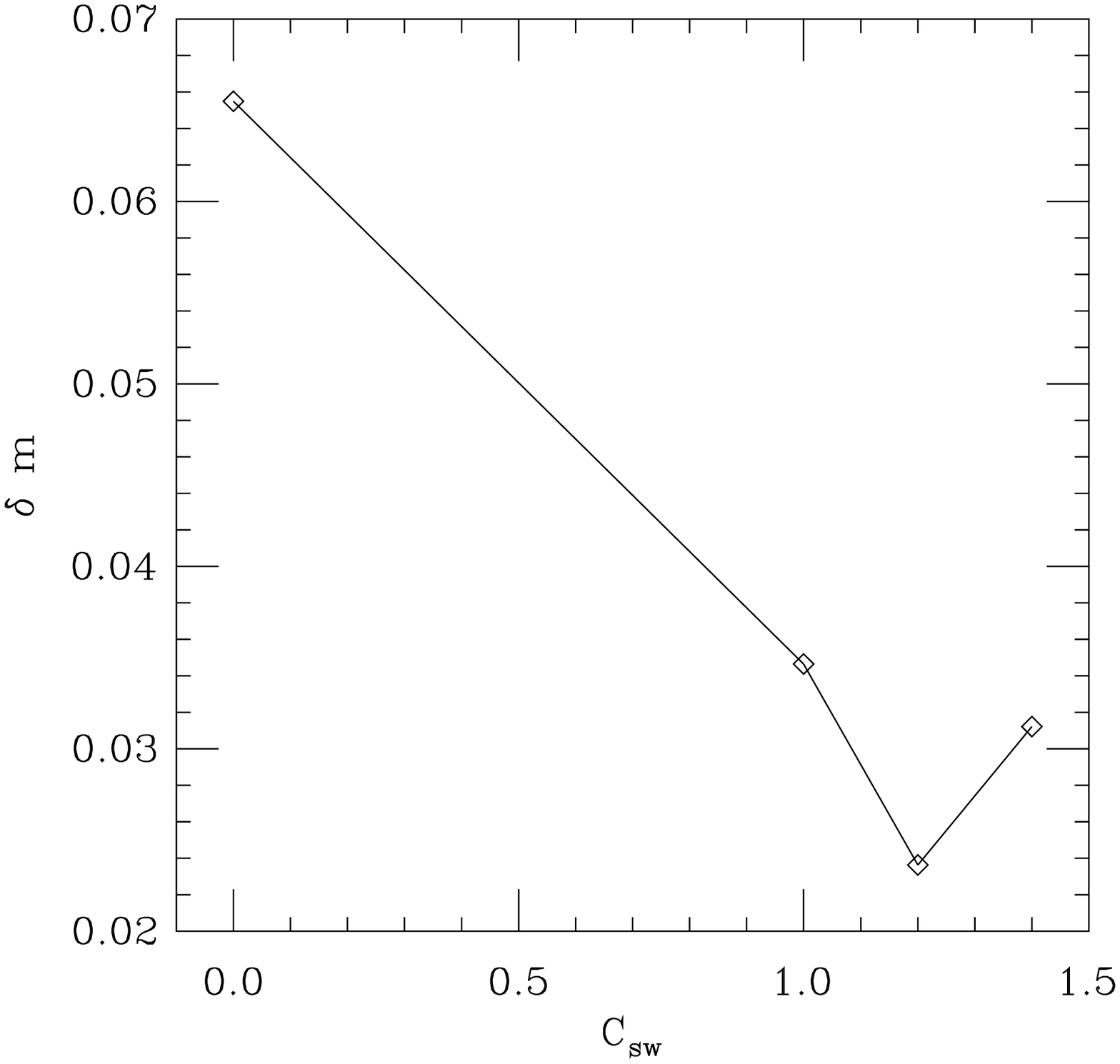}{80mm}}
\caption{(a) The average real eigenmode location as the 
function of the clover coefficient at $\beta=5.8$. The error
bars show the spread of the modes. (b) The spread of 
the modes from (a) as the function of
the clover coefficient. }
\label{fig:averpole}
\end{figure}

To quantify the spread of the low energy real modes, 
in Fig. \ref{fig:averpole}a we plot the average real eigenmode
location as the function of the clover coefficient. 
The error bars here are not errors, they are the
spread of the modes in $m$.  
Fig. \ref{fig:averpole}b shows only the spread as the function of the
clover coefficient.
$m_c$, where the pion becomes massless, lies somewhere in the middle of
the mode distribution, closer to its top (large mass)
end. The average eigenmode locations
and the upper end of the error bars in Fig. \ref{fig:averpole}a
bracket $m_c$.
Smearing the link removes most of the additive mass renormalization even for
the Wilson action, changing $m_c \sim -0.95$ for the thin link action 
to $m_c \sim -0.22$ for our case. 
Adding a clover term to the action further reduces the
additive mass renormalization, and even larger clover terms
induce a positive mass renormalization. At $\beta=5.8$ the additive mass 
renormalization is minimal for $c\approx 1.2$. 
This is also the value where the spread of the eigenmodes is minimal.
To conclude, the results indicate that at $\beta=5.8$
 (lattice spacing $a\simeq 0.15$ fm) the low lying  and
doubler modes are well separated and explicit 
chiral symmetry breaking is minimized with clover
coefficient $C_{sw}=1.2$ with our fat link action.

\begin{figure}
\centerline{\ewxy{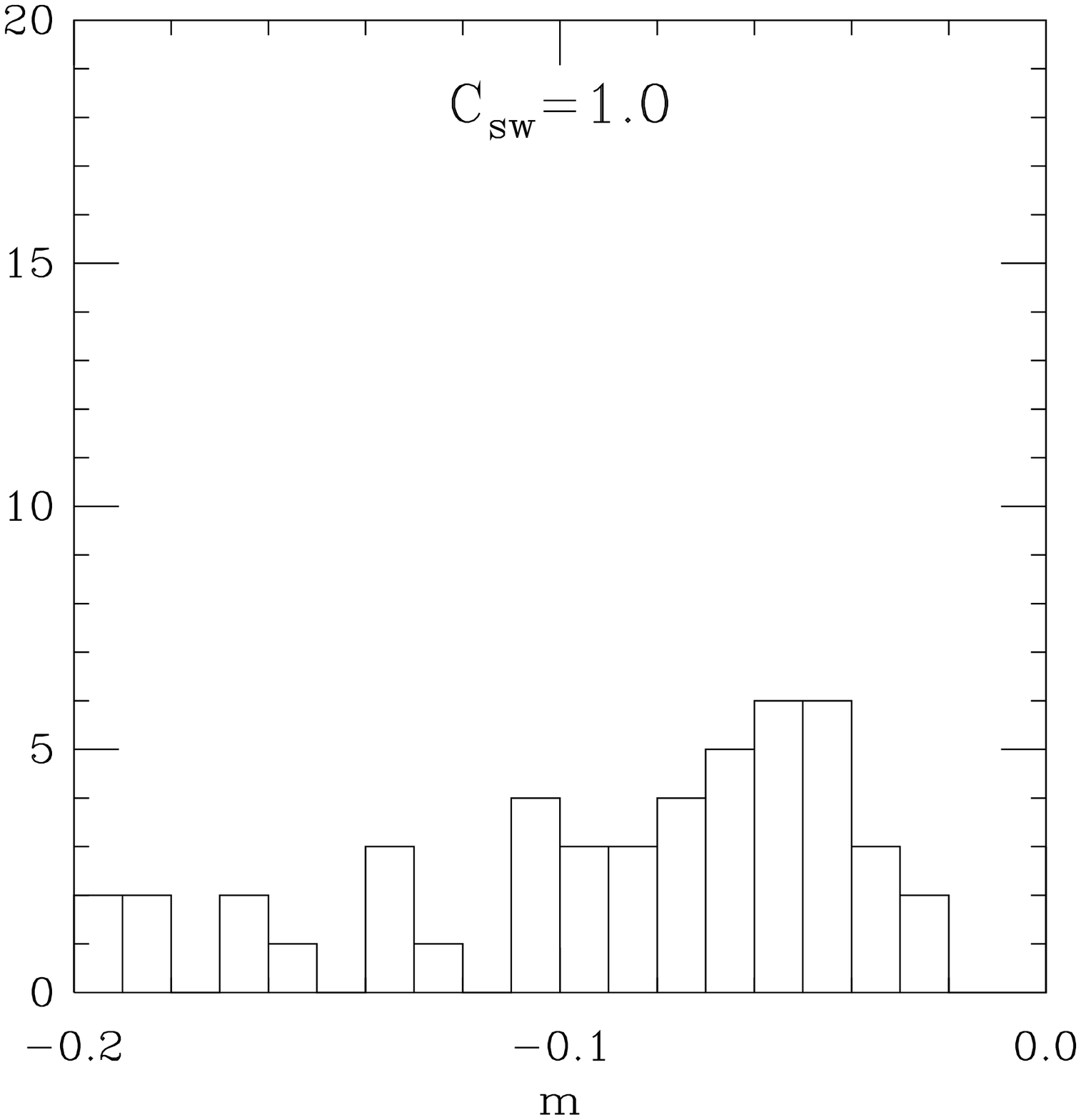}{80mm}
\ewxy{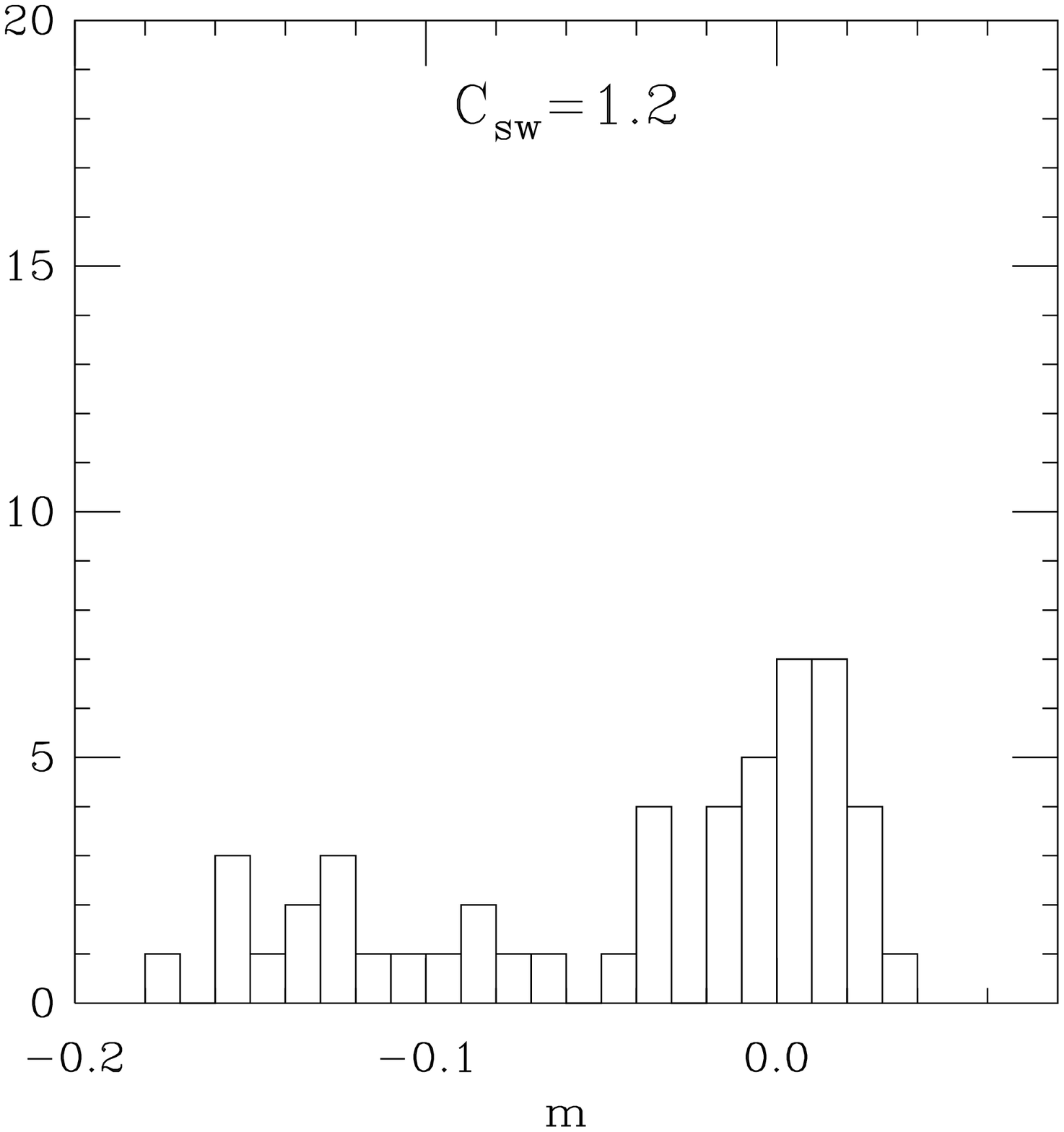}{80mm}}
\vspace{0.5cm}
\centerline{\ewxy{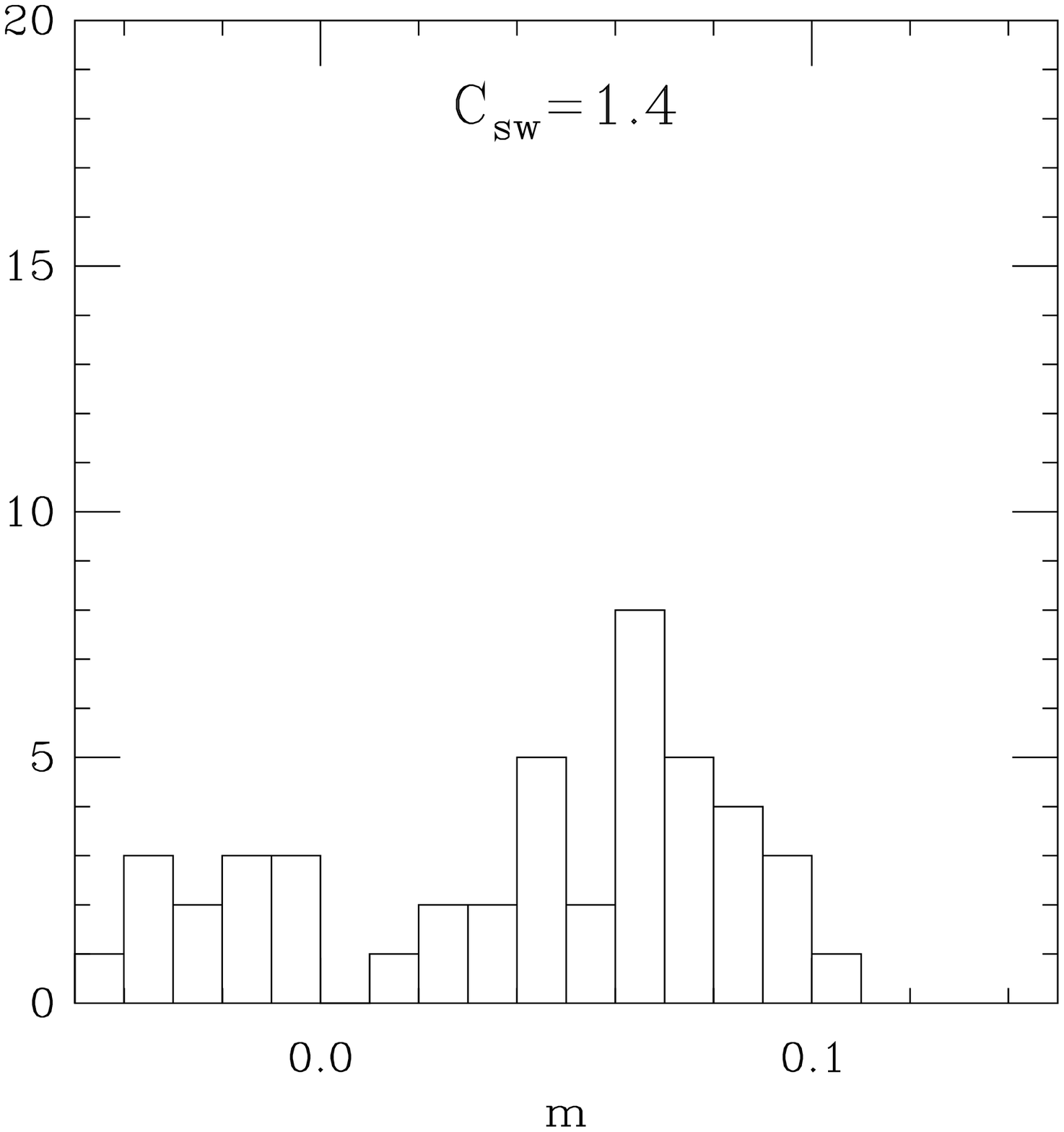}{80mm}}
\caption{The distribution of the bare masses $m$, where the 
quark propagator is singular at $\beta=5.7$  of fat link  clover
fermions $C_{sw}=1.0$ (a);  $C_{sw}=1.2$ (b); $C_{sw}=1.4$ (c).}
\label{fig:polevsclover5.7}
\end{figure}

The situation is less convincing at $\beta=5.7$ (lattice spacing 
$a\simeq 0.20$ fm). Fig.
\ref{fig:polevsclover5.7}, again based on 40 $8^4$ configurations,
 shows the distribution of the eigenmodes at $\beta=5.7$, clover coefficient
$C_{sw}=1.0,1.2$ and 1.4. Here the low lying  modes are not as 
well separated as for $\beta=5.8$. For
$C_{sw}=1.0$ the distribution has a large tail 
extending towards negative quark masses. The situation is a bit
better for $C_{sw}=1.2$. Since the low lying modes are not well
separated from the doublers, it makes no sense to calculate 
the spread of the distribution. The
continuum description of chiral symmetry breaking 
is about to break down at $\beta=5.7$. If any of the distributions of Fig. 
\ref{fig:polevsclover5.7} describes continuum physics, it is $C_{sw}=1.2$ where the
overall mass renormalization is close to zero and 
the low lying modes are best separated form the
doublers. 

\begin{figure}
\centerline{\ewxy{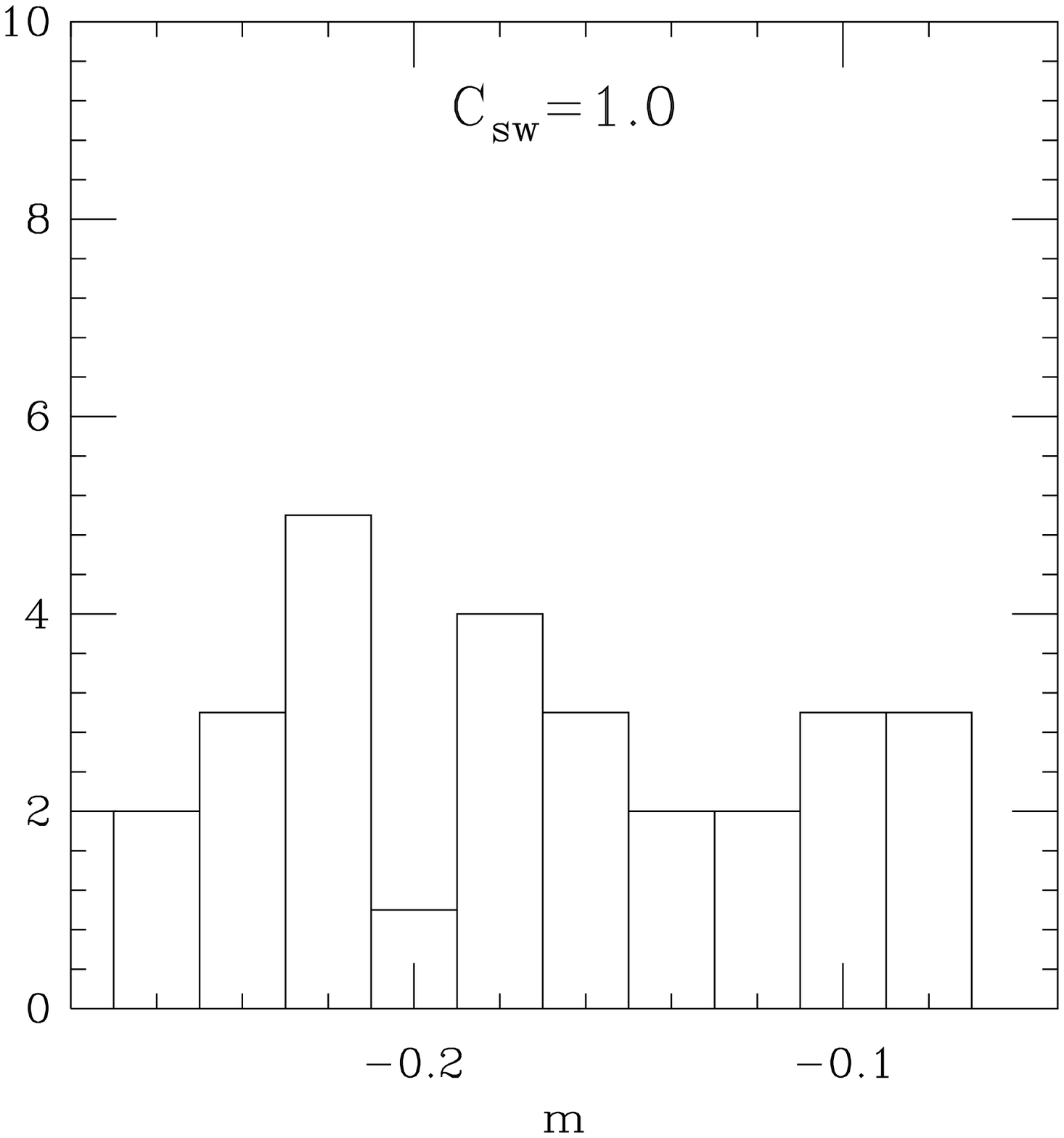}{80mm}
\ewxy{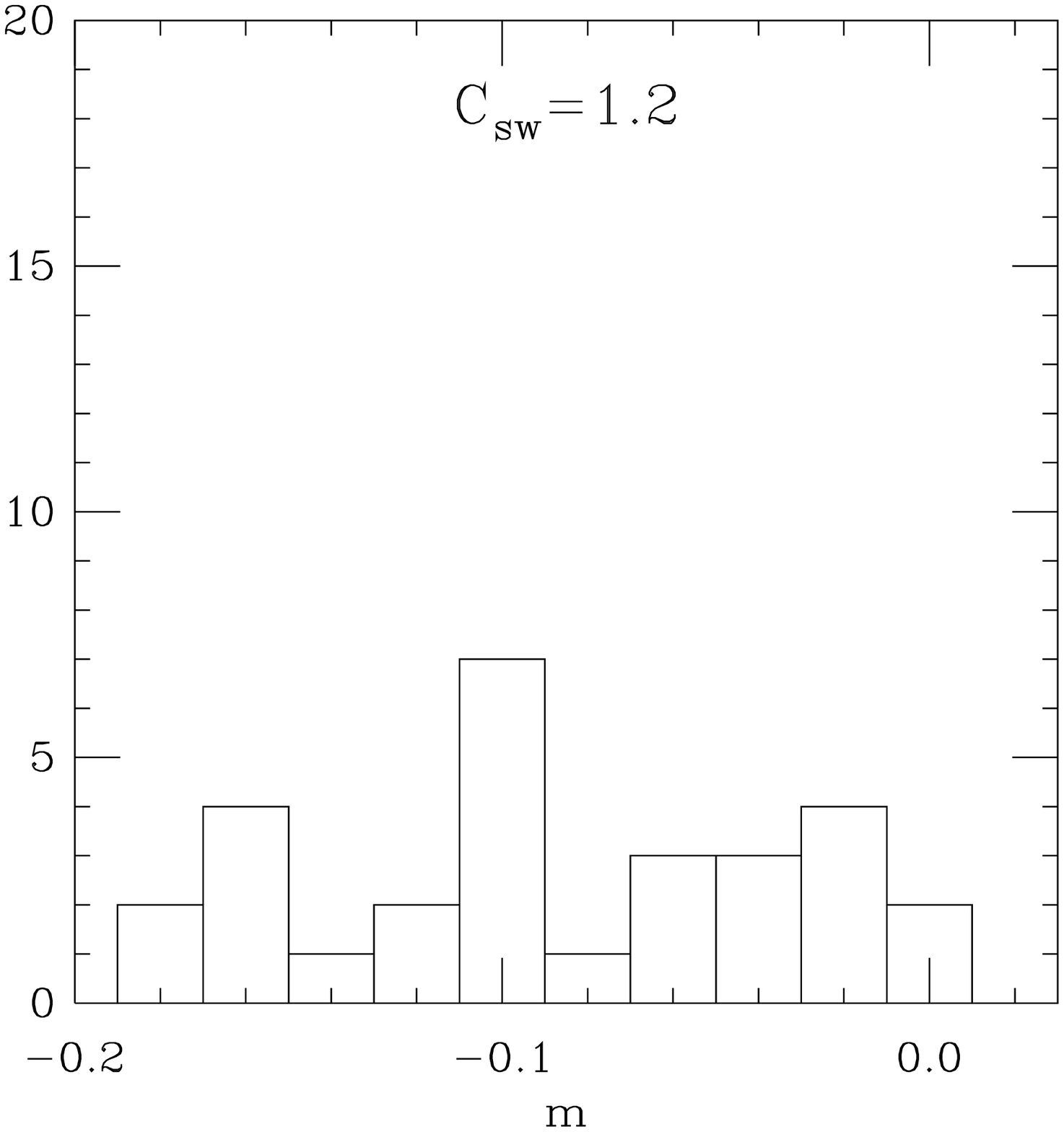}{80mm}}
\caption{The distribution of the bare masses $m$, where the 
quark propagator is singular at $\beta=5.55$  of fat link  clover
fermions $C_{sw}=1.0$ (a);  $C_{sw}=1.2$ (b).}
\label{fig:polevsclover5.55}
\end{figure}

Finally, at $\beta=5.55$ (lattice spacing $a\simeq 0.24$ fm)
Fig. \ref{fig:polevsclover5.55} shows that the distribution
of eigenmodes is broad and the low lying  modes and doublers are completely
mixed.

These figures show that it is not possible to make the lattice spacing
greater than about 0.2 fm, and still retain the continuum-like description
of chiral symmetry breaking using a clover-like action. 
Other, better tuned actions might perform
better at large lattice spacing. Our exploratory studies with a 
hypercubic fixed point \cite{HYPER} action showed
that the FP action is not better in terms of chiral 
symmetry breaking than the fat link clover action
with $C_{sw}=1.0$. Its dispersion relation was improved
at lattice spacing $a=0.36$ fm, and it (and the clover action)
showed only a small amount of scale violation in hyperfine splittings
at large lattice spacing.
This just shows that scaling or near scaling of a 
few quantities does not guarantee that the
physical mechanism responsible
for  chiral symmetry breaking is the same as in the continuum.
 This should serve as a warning sign for any calculations
at large lattice spacing with actions of untested chiral properties.

In principle, one should optimize the spread not in terms 
of the bare quark mass, but in terms of some corresponding 
physical observable, like the pion mass. 
However, at large fattening, the relation between the
bare quark mass and the pion mass shows little variation with $C_{sw}$.

The usual nonperturbative tuning of the clover term \cite{LUSCHER}
attempts to optimize current algebra via PCAC relations.
It is not clear to us how this is related to what we do, since
it has been done so far only for thin link actions.  We know that
the optimal $C_{sw}$ by our criterion for a thin link action is
larger than the usual nonperturbative tuning, for example 2.5 vs 2.0   at
$\beta=5.8$.  Our criterion is based on a direct attack on low energy real
eigenmodes.
To the extent that the
dynamics of QCD for light quark masses is 
dominated by this physics, we feel our
optimization criterion is well founded.

We do not know if there is a connection between this tuning method
and the Ginsparg-Wilson \cite{GW} realization of chiral symmetry on the
lattice. The free field limit of these fat link actions is identical
to the standard Wilson action, and so while  we are 
 improving the
chiral properties of the interacting theory, we did not improve
 the chiral properties
of the free theory.

\section{Numerical Tests}
We have done a calculation of spectroscopy and matrix elements
at lattice spacing $aT_c=1/4$ 
($\beta=5.7$ with the Wilson gauge action)
using the fat link clover action with $C_{sw}=1.2$,
on a set of 80 $8^3\times24$ Wilson $\beta=5.7$
configurations.
This is on the edge of the allowed range of lattice spacings according
to our criterion of the last section.
By itself, one lattice spacing is not a scaling test, but we can combine
our results with those from other actions to compare the new action
to them.

\begin{figure}[h!tb]
\begin{center}
\vskip 10mm
\leavevmode
\epsfxsize=100mm
\epsfbox{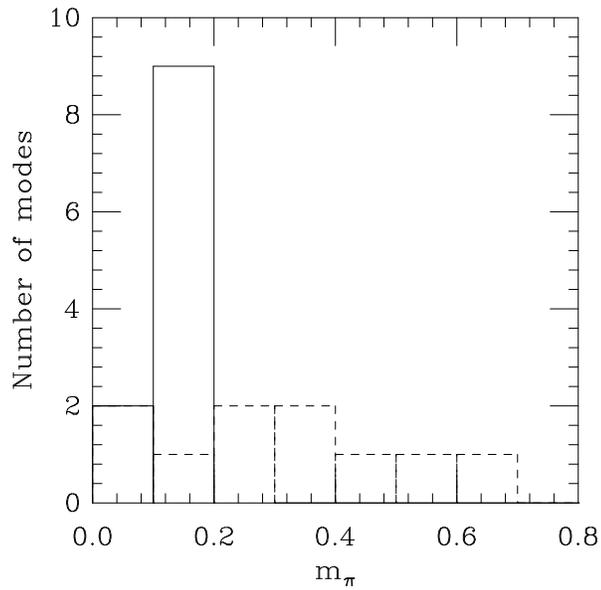}
\vskip 10mm
\end{center}
\caption{The distribution of exceptional fermionic modes
for the thin link clover action with the non-perturbatively
determined value of $C_{sw}=2.25$ (dashed line)
and the optimized fat link action (full line). The eigenvalues
are quoted in terms of the pion mass (in lattice units) where
the corresponding eigenmode makes the quark propagator singular.}
\label{fig:pole_dist}
\end{figure}

We expect that minimizing the
spread of real eigenvalues will also improve the situation
with exceptional configurations. This is indeed what happens.
As a preliminary calculation, we determined all the exceptional modes
on a set of 40 $6^3\times16$ Wilson $\beta=5.7$ configurations
both for our optimized
fat link action and the thin link clover action with $C_{sw}=2.25$.
The histograms of Fig.\ \ref{fig:pole_dist} show
where the exceptional modes occur in terms of the lattice pion masses
for both actions.
For the fat link action all the modes occurred at pion masses
$m_{\pi}<0.3$ which corresponds to $m_{\pi}/m_{\rho} \leq 0.38$,
whereas the thin link clover action had five exceptional
modes above $m_{\pi}=0.3$ at $m_{\pi}= 0.42, 0.53, 0.39, 0.30,
0.62$ (in lattice units).

On the $8^3\times 24$ spectroscopy data set
 we only identified the eigenmodes occurring above
$m_\pi=0.3$, as going any closer to the critical mass would
have been prohibitively costly. With the thin link $C_{sw}=2.25$
action we found 11 poles above $m_\pi=0.3$, the most outlying
being at $m_\pi=0.7$. Using the fat link action no pole was found
above $m_\pi=0.3$.

We also noticed that the new action seems to be more convergent than 
the thin link clover action: the same biconjugate gradient code needs about
half as many steps to converge to the same residue as the usual thin link 
clover action, for the same $\pi/\rho$ mass ratio.

\subsection{Spectroscopy}

The spectroscopy measurement is entirely straightforward. 
We gauge
fixed to Coulomb gauge and used a Gaussian independent particle source
wave function $\psi(r) =
\exp(-(r/r_0)^2)$ with $r_0=2$.
We used pointlike sinks projected onto low momentum states.
We used naive currents ($\bar \psi \gamma_5 \psi$, etc.)
 for interpolating fields.
The spectra appeared to be asymptotic (as shown by good (correlated)
 fits to a single exponential)
beginning at $t\simeq 3-5$  and the best fits were selected using the 
 HEMCGC criterion \cite{HEMCGC}.

Our fiducials for comparison
are Wilson action and clover action
quenched spectroscopy.  We have tried to restrict the data
we used for comparison to lattices with the proper physical volume.

\begin{figure}
\centerline{\ewxy{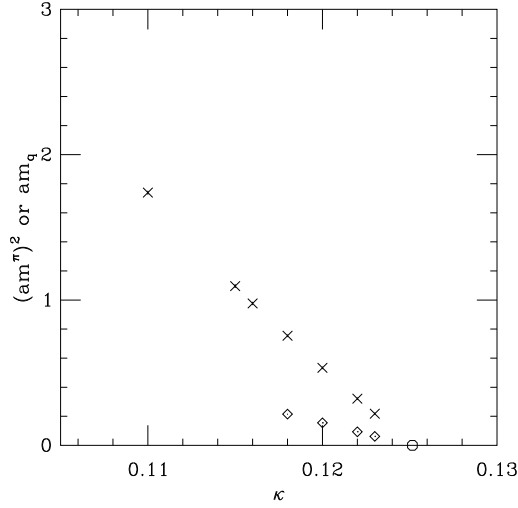}{80mm}
}
\caption{Bare squared pion mass  (crosses) and
quark mass from Eq. 12 (diamonds) 
 vs hopping parameter for the $C_{sw}=1.2$ action.
The octagon shows $\kappa_c$.}
\label{fig:pisqc1.2}
\end{figure}

We can roughly estimate the critical bare quark
mass (at which the pion is massless)
by  linearly extrapolating $m_\pi^2$ to zero in $m_0$.
We also estimated the critical bare mass using the PCAC relation
\bee
\nabla_\mu \cdot \langle \bar\psi \gamma_5 \psi(0) \bar\psi\gamma_5 \gamma_\mu
\psi(x) \rangle =
2m_q \langle\bar\psi \gamma_5 \psi(0) \bar\psi\gamma_5 \psi(x) \rangle .
\ee
which, going to the lattice and following \cite{DOUGFPI},
is done by fitting the pseudoscalar source-pseudoscalar sink to
\bee
P(t) = Z(\exp(-m_\pi t) + \exp(-m_\pi (N_t -t) ) )
\ee
and the pseudoscalar source-axial sink to
\bee
A(t) = {Z_P \over Z_A}
{{2m_q}\over m_\pi} Z(\exp(-(m_\pi t) - \exp(-m_\pi (N_t -t) ))
\ee
to extract $m_q$.
Fig. \ref{fig:pisqc1.2} shows the
squared pion mass vs bare quark mass. 
The quark masses from the local pseudoscalar and axial currents
are also shown.

Fits of the squared pion mass $(am_\pi)^2 = B(1/\kappa -1/\kappa_c)$
give $B=1.55(2)$ and $\kappa_c=0.12515(7)$.
The quark mass is fit to $am_q = A(1/\kappa -1/\kappa_c)$
and we find $A=0.447(7)$, $\kappa_c=0.12515(6)$. In free field theory,
we would expect $A=1/2$, $\kappa_c=1/8$,
 and this is our first hint that perturbative corrections
to bare quantities are small.

\begin{figure}
\centerline{\ewxy{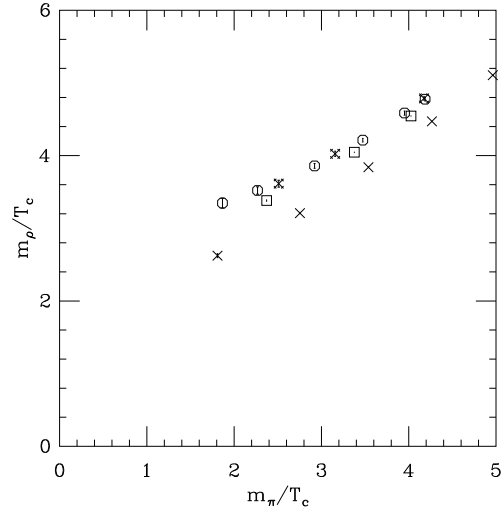}{80mm}
}
\caption{Octagons show $m_\rho/T_c$ vs. $m_\pi/T_c$ for  the $C_{sw}=1.2$ action
at $aT_c=1/4$. Also shown are Wilson action data, with
crosses for $aT_c=1/4$,
squares for $aT_c=1/8$,
and
fancy crosses for $aT_c=1/12$.
}
\label{fig:rhotccl}
\end{figure}

\begin{figure}
\centerline{\ewxy{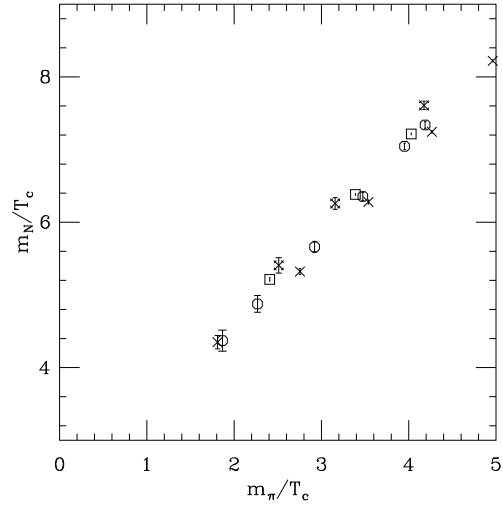}{80mm}
}
\caption{$m_N/T_c$ vs. $m_\pi/T_c$ for  the $C_{sw}=1.2$ action
and the Wilson action, labeled as in Fig. 6.}
\label{fig:ntccl}
\end{figure}

As a scaling test we compare
$m_\rho/T_c$ and $m_N/T_c$ vs. $m_\pi/T_c$ for the $C_{sw}=1.2$ action in
 Figs.  \ref{fig:rhotccl}
and  \ref{fig:ntccl}.

\begin{figure}
\centerline{\ewxy{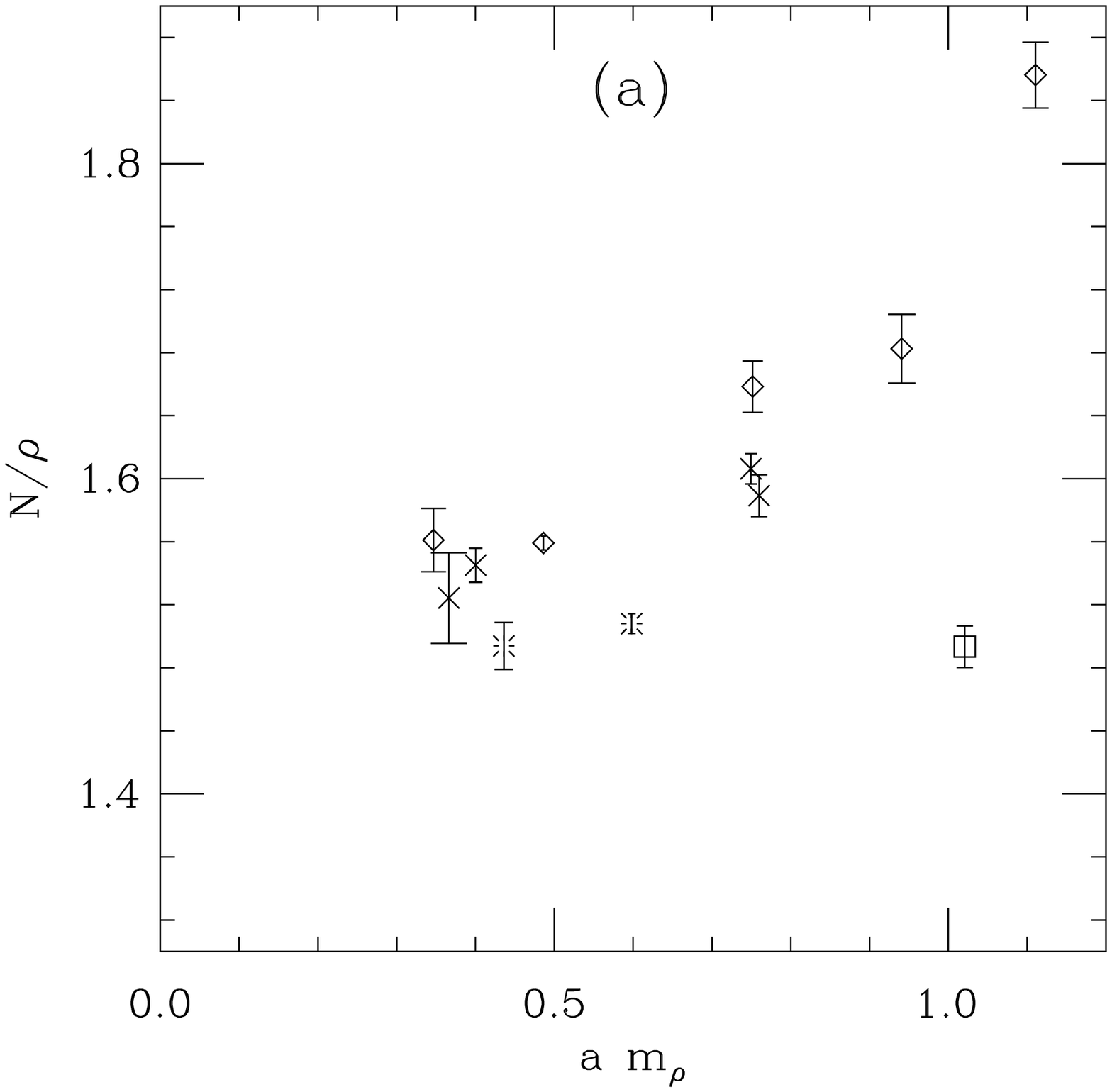}{80mm}
\ewxy{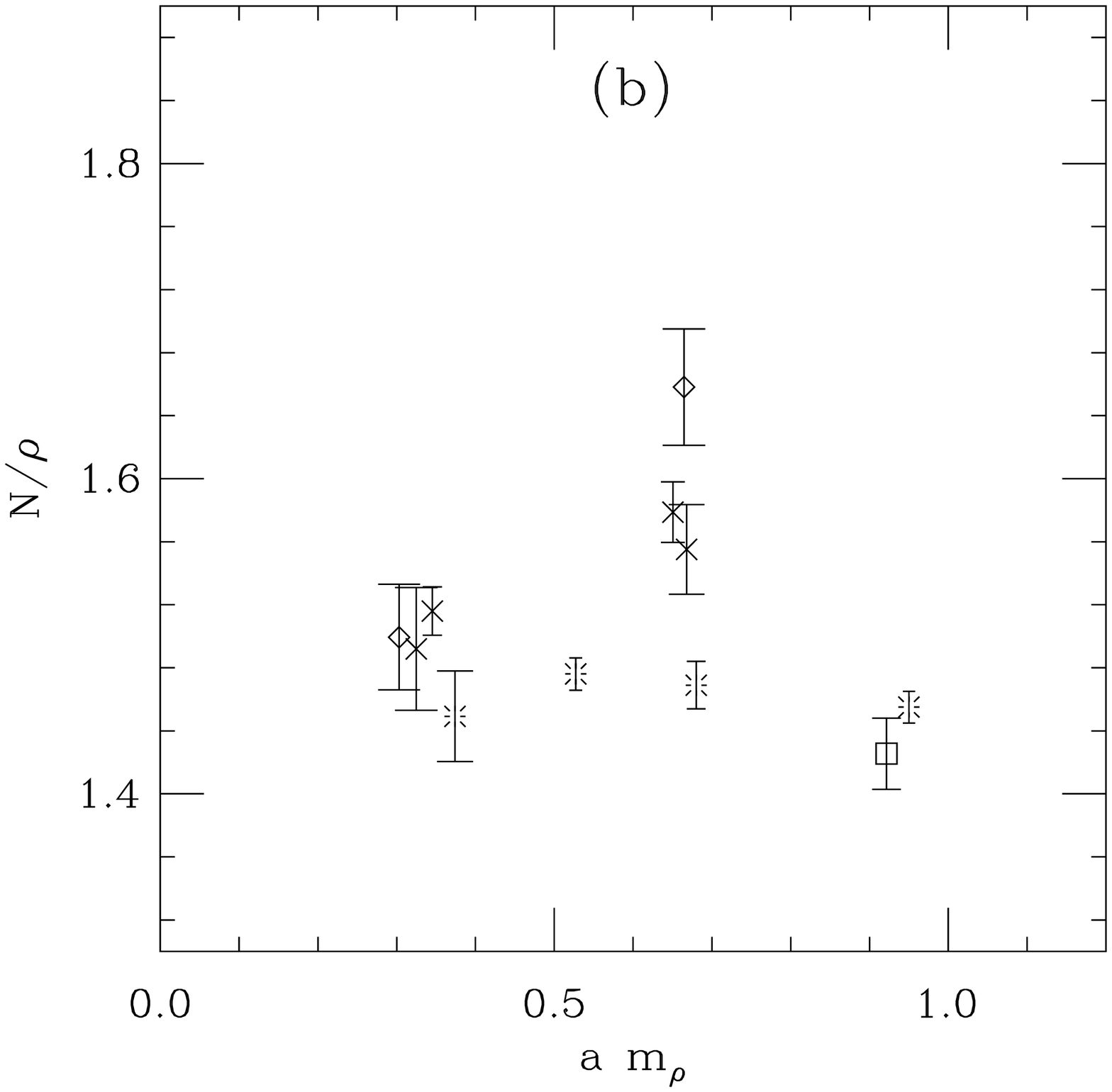}{80mm}}
\vspace{0.5cm}
\centerline{\ewxy{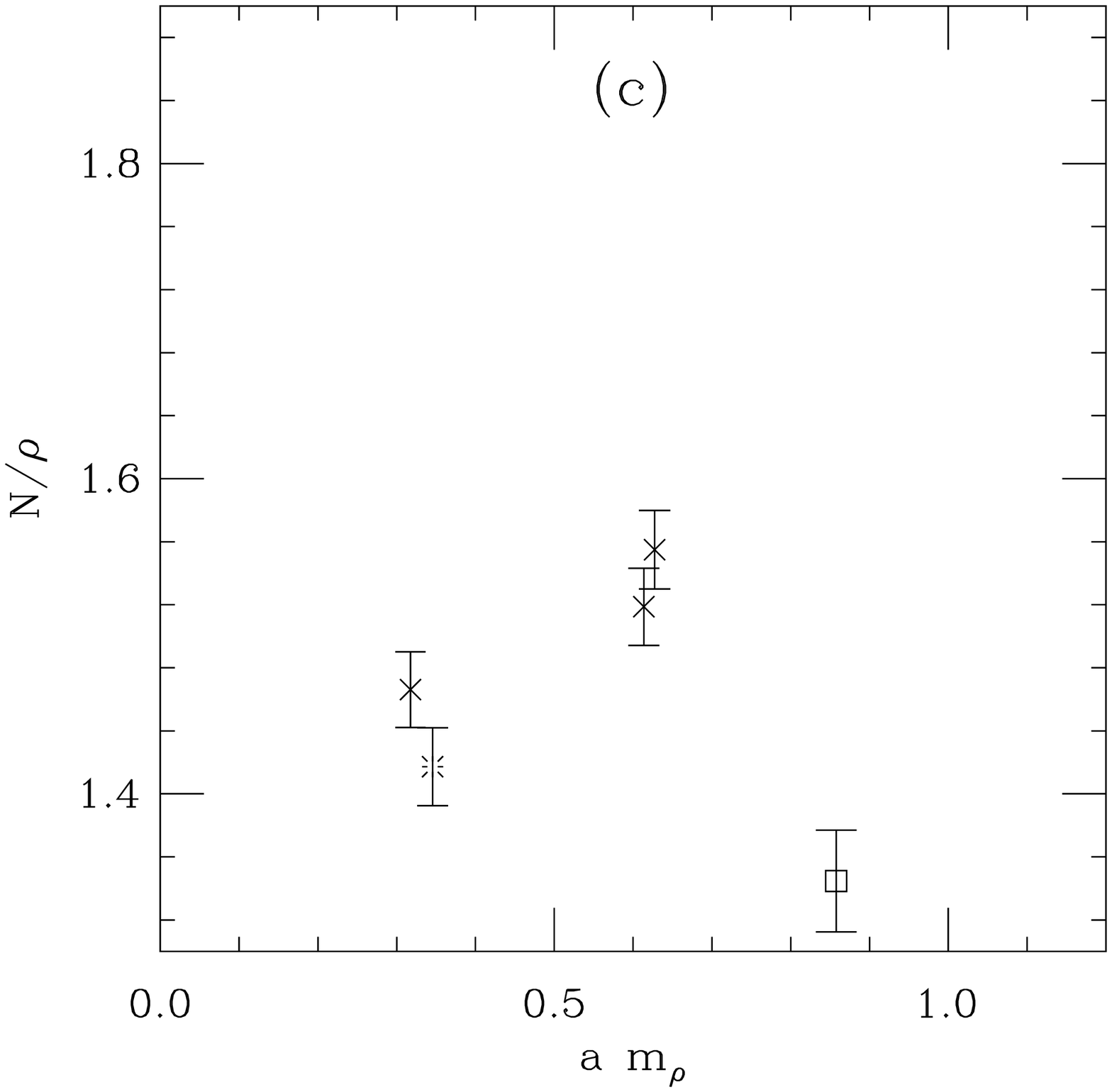}{80mm}}
\caption{A scaling test for the $C_{sw}=1.2$ action (square)
vs. Wilson actions on lattices of fixed physical size (diamonds)
and larger volumes (crosses), and the
nonperturbatively improved clover action (bursts).
 Data are interpolated to
$\pi/\rho=0.80$ (a), 0.70 (b), and 0.60 (c).}
\label{fig:allrat}
\end{figure}

We  compare scaling violations in hyperfine splittings by
interpolating our data to fixed $\pi/\rho$ mass ratios and plotting
the $N/\rho$ mass ratio vs. $m_\rho a$. We do this at three
$\pi/\rho$ mass ratios, 0.80, 0.70 and 0.60, in Fig. \ref{fig:allrat}.
In these figures the diamonds are Wilson action data in lattices of fixed
physical size,
$8^3$ at $\beta=5.7$ \cite{BUTLER},
$16^3$ at $\beta=6.0$ \cite{DESY96173}
$24^3$ at $\beta=6.3$ \cite{APE63},
 and the crosses are data in various larger lattices:
$16^3$ and $24^3$ at $\beta=5.7$ and $32^3$ at $\beta=6.17$ \cite{BUTLER},
$24^3$ at $\beta=6.0$ \cite{DESY96173}.
When they are present the data points from larger lattices illustrate
the danger of performing scaling tests with data from different volumes.
The bursts are from the nonperturbatively improved clover action of Refs.
\cite{ALPHA} and \cite{SCRI}.
The squares show the $C_{sw}=1.2$ action.

We give a list of masses from the $C_{sw}=1.2$ action in Table
 \ref{tab:w570}.

We conclude that the fat link clover action has at least as good
scaling behavior as the usual nonperturbatively tuned clover action.

\subsection{Dispersion Relations}

\begin{figure}[htb]
\begin{center}
\vskip -10mm
\leavevmode
\epsfxsize=60mm
\epsfbox[40 50 530 590]{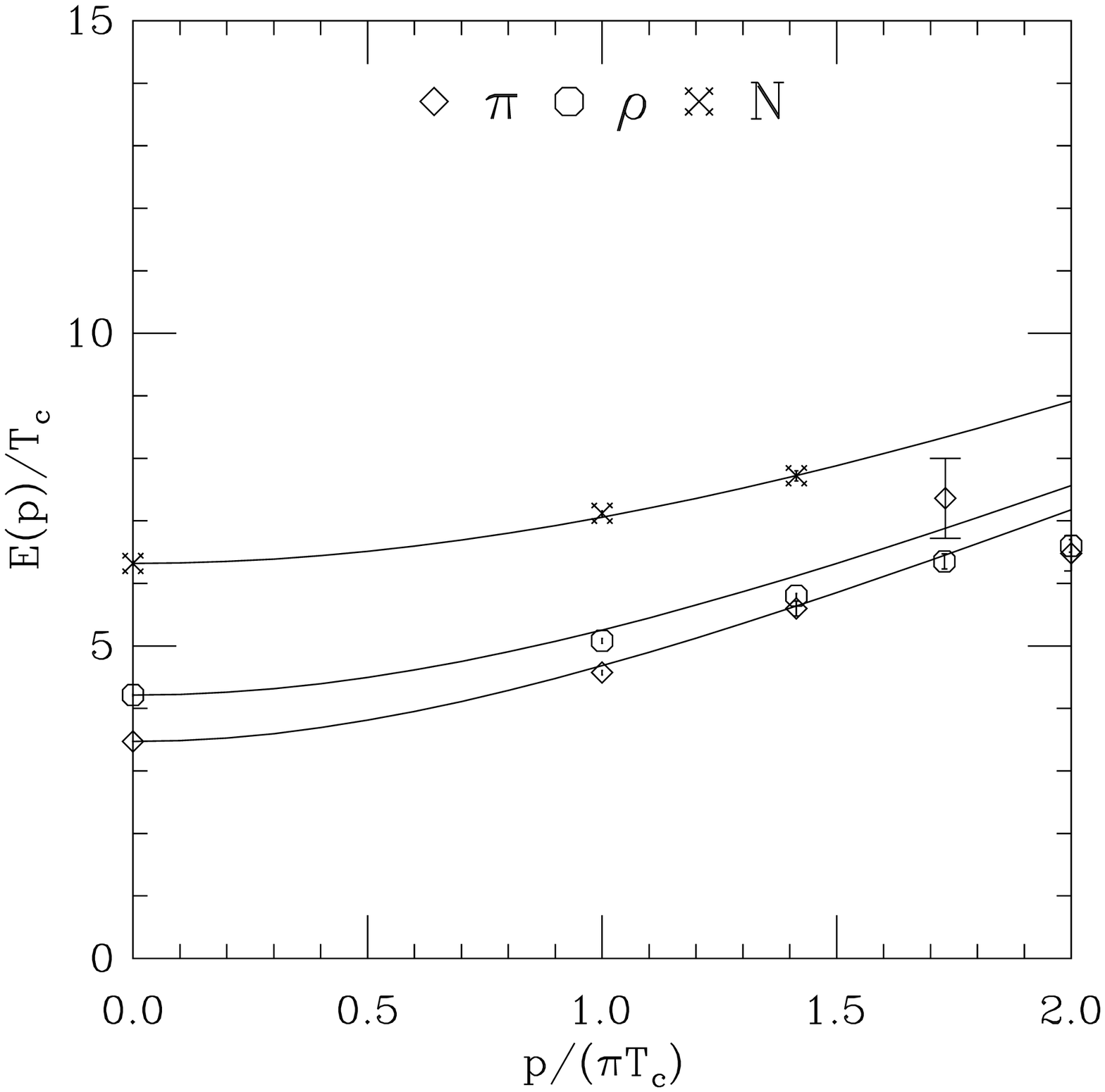}
\vskip -5mm
\end{center}
\caption{Dispersion relation for  hadrons at $aT_c=1/4$ ($a \simeq 0.18$
fm)  from the $C_{sw}=1.2$ action.
The curves are the continuum dispersion relation for the appropriate
(measured) hadron mass.}
\label{fig:eptc}
\end{figure}

To view the dispersion relation, we first
plot $E(p)$, the energy of the state produced with spatial momentum
$\vec p$, as a function of $|\vec p|$.
The result for   the $C_{sw}=1.2$ action  at $\kappa=0.118$ is compared to the
free dispersion relation 
at $aT_c=1/4$ in Fig. \ref{fig:eptc}.
It looks very similar to results from the Wilson action at the same
lattice spacing--an entirely unsurprising result.

\begin{figure}
\centerline{\ewxy{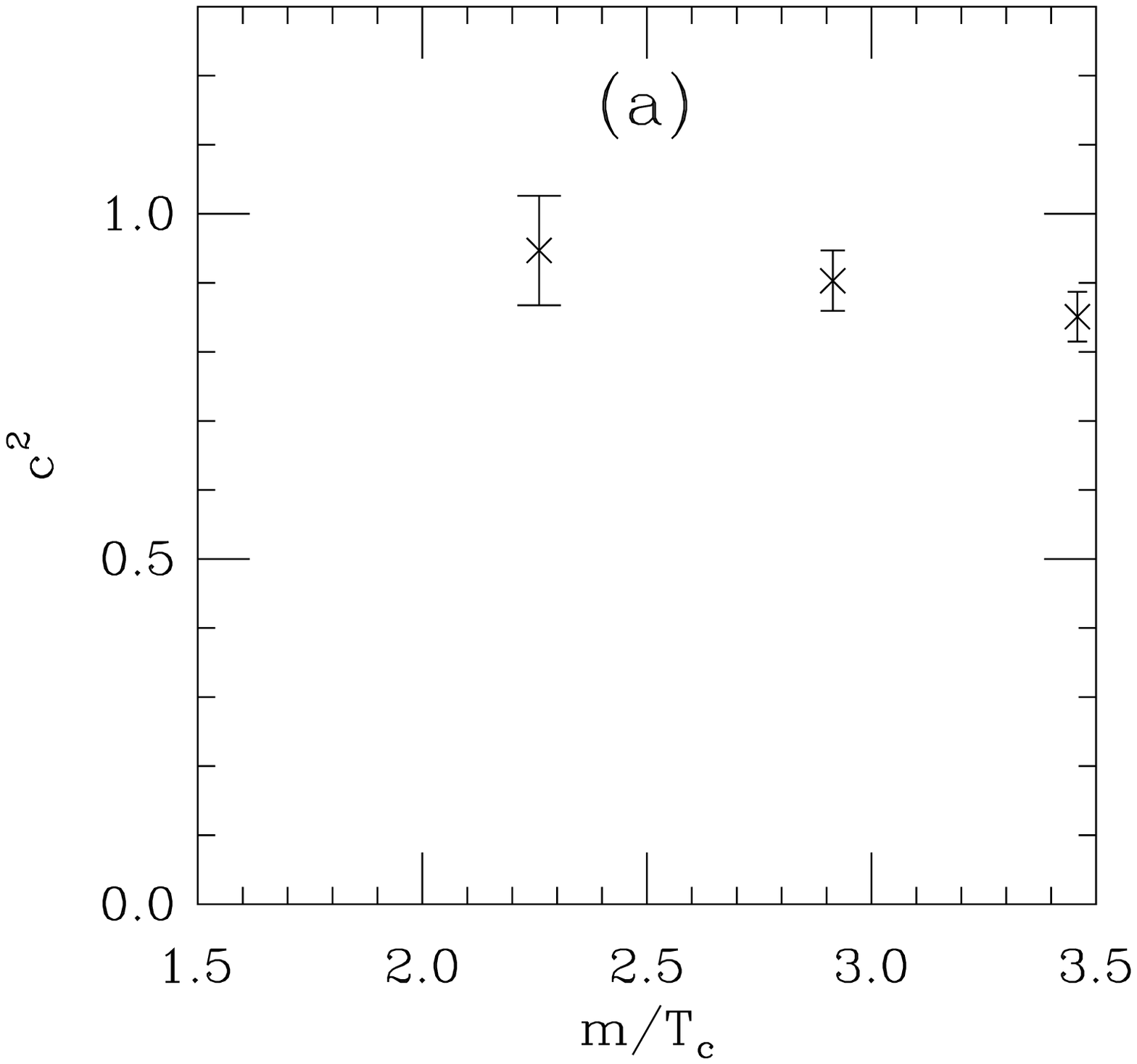}{80mm}
\ewxy{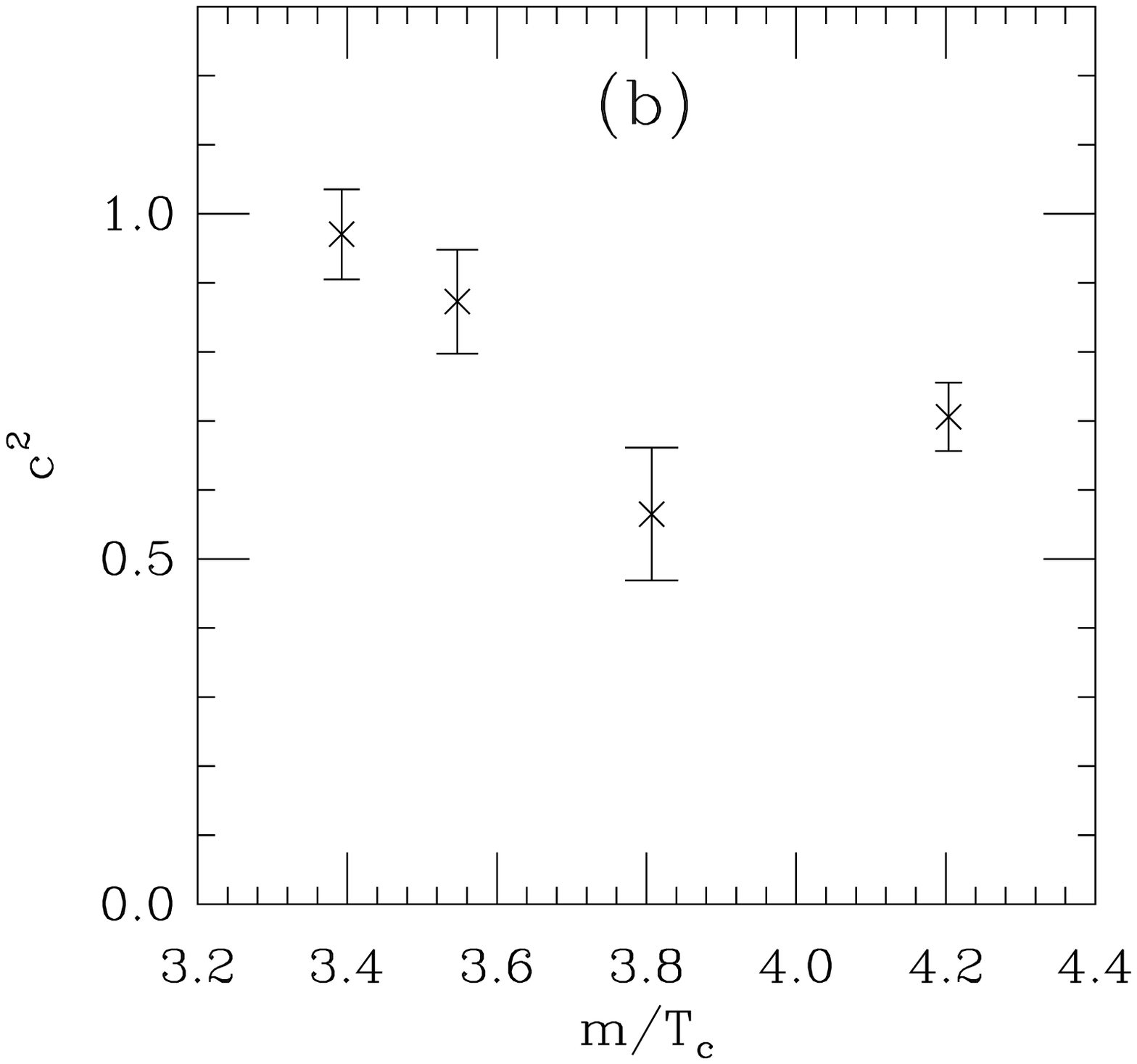}{80mm}}
\vspace{0.5cm}
\centerline{\ewxy{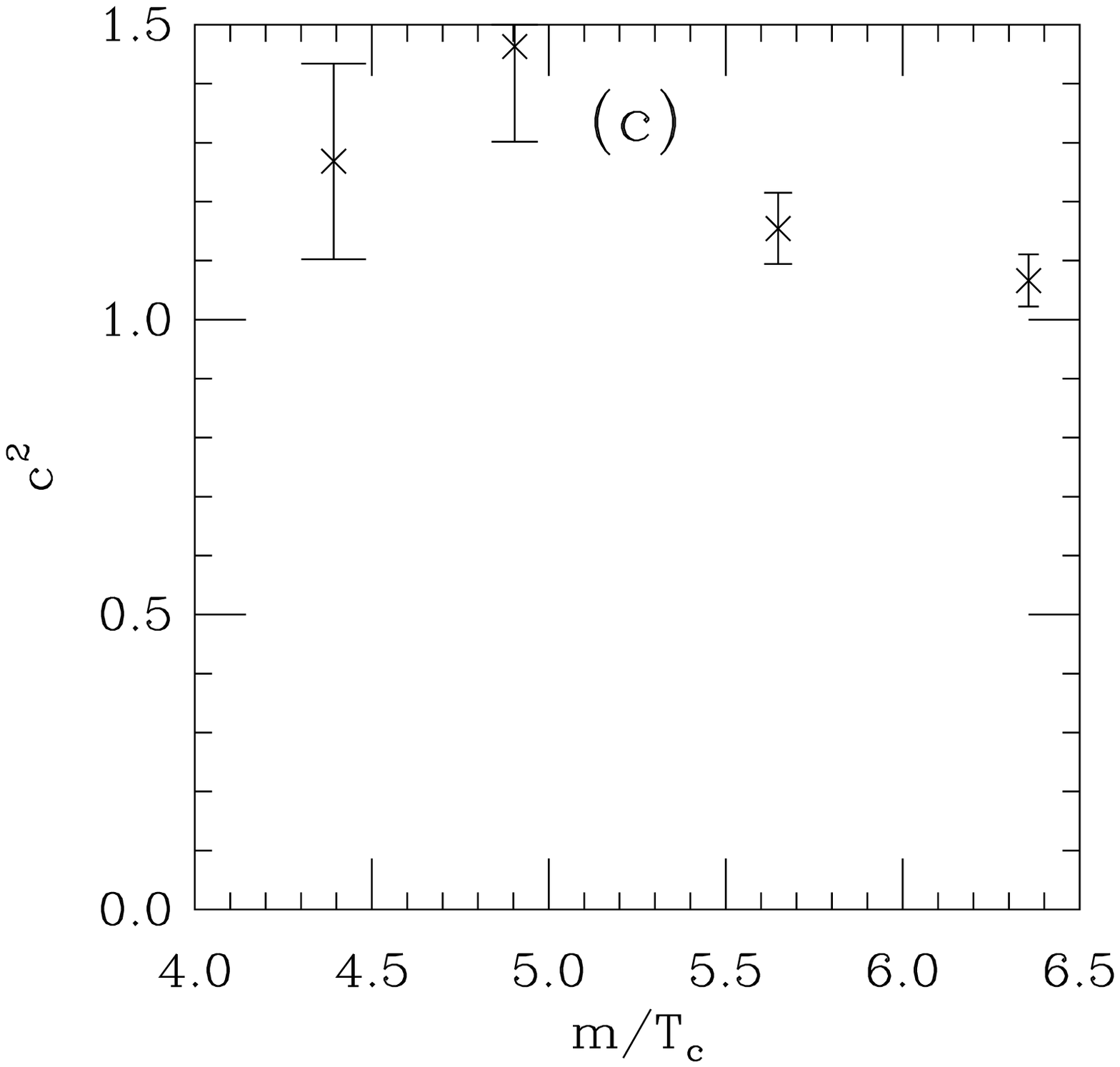}{80mm}}
\caption{Squared speed of light vs. hadron mass in units of $T_c$,
for (a) pseudoscalars, (b) vectors) and (c) protons, from the
$C_{sw}=1.2$ action at $\beta=5.7$.}
\label{fig:csq}
\end{figure}

This behavior is quantified by measuring the squared speed of light,
$c^2 = (E(p)^2-m^2)/p^2$, for $\vec p= (1,0,0)$.  We do this by performing
a correlated fit to the two propagators. The result is presented in 
Fig. \ref{fig:csq} and shows the worsening of the dispersion relation
at larger quark mass, characteristic  Wilson/clover kinetic behavior.

\begin{table}
\begin{tabular}{|c|l|l|l|l|l|}
\hline
$\kappa$ & PS  & V  &  N  &  $\Delta$ \\
\hline
   0.118 &  0.868( 3) &  1.053( 6)&  1.589(13) &  1.725(19)  \\
   0.120 &  0.730( 4) &  0.965( 8)&  1.415(19) &  1.593(22)  \\
   0.122 &  0.567( 5) &  0.880(12)&  1.219(28) &  1.501(24)  \\
   0.123 &  0.467( 6) &  0.837(17)&  1.093(36) &  1.432(27)  \\
\hline
\end{tabular}
\caption{Table of best-fit masses, $C_{sw}=1.2$, $\beta=5.7$.}
\label{tab:w570}
\end{table}

\subsection{Renormalization factors}
We now turn to a (rather naive) set of measurements of simple matrix
elements.

Formally, the fat link action can be regarded as ``just another''
 $O(a^2)$ improved action, if one assumes that the coefficient of
the clover term is actually $C_{sw}= 1+ O(g^2)$.  In perturbation theory,
which involves the vector potential $A_\mu$ rather than the link,
the action shows the usual cancellation between the clover term
and the scalar $\bar \psi \psi A$ term. The spectrum is $O(a^2)$;
matrix elements using ``rotated fields'' $\psi \rightarrow
(1 -(\gamma\cdot D )/2) \psi$ are $O(a^2)$ improved.  Of course, the 
lattice-to-continuum 
renormalization Z-factors
are different than the usual thin link clover Z-factors.

In principle, one could calculate all 
 Z-factors using lattice perturbation theory. The Feynman
rules differ from the rules for the usual clover action, only in
 that the vertices are multiplied by form factors, which are
functions of the gluonic momentum.  We have not tried to do this yet.
Since we are working at fairly strong coupling, we felt that it would
be easier to compute the Z-factors nonperturbatively, beginning with
the vector current and then computing the axial 
renormalization factor using Ward identities.  Since this is just the
first calculation using this action, we will restrict ourselves to
the naive (local) currents.

We begin with the vector current.  The conserved (Noether) current
for the fat link action is the same as the Wilson current, just with
fat links instead of thin links,
\bee 
J_\mu^{cons}(n) = {1 \over 2} (\bar \psi(n)(\gamma_\mu -1)V_\mu(n)\psi(n+\hat \mu)
- \bar \psi(n+\hat \mu)(\gamma_\mu +1)V_\mu^\dagger(n)\psi(n) ).
\ee
This current is conserved but not improved.  We choose to define the 
vector current Z-factor from the ratio of  forward matrix elements
\bee
Z_V = {{\langle \pi(T) \pi(0) \rangle} \over
 {\langle \pi(T) J_0(t)\pi(0) \rangle}}
\label{ZV}
\ee
for $0<t<T$.
The local current is defined as $J_\mu^{loc}(n) = \bar \psi(n) \gamma_\mu \psi(n)$.
We measured the Z-factors using 20 configurations on a periodic
 $8^3 \times 24$
lattice with $T=10$ and averaged $9>t>1$. The results of a jacknife analysis
are shown in Table \ref{tab:zv}.

The fact that $Z^{cons} \ne 1$ is a finite size effect.  In the numerator
of Eqn. \ref{ZV} both quarks can propagate in  both time directions around
the torus, but the forwards and backwards paths contribute differently
to the denominator. We correct this phenomenologically by computing the
ratio $Z^{loc}/Z^{cons}$ and display it in the table.

Lattice perturbation theory which correctly takes into account the residue
of the pole of the massive quark\cite{ZFACTOR}  predicts that this ratio is
equal to 
\bee
Z_V= {{1-6 \kappa}\over{2\kappa}} \hat Z_V
\ee
 where the $(1-6 \kappa)$ factor is the
residue (recall $\kappa_c=.125$) and $\hat Z_V = 1 + a g^2 +\dots$. 
 As Table \ref{tab:zv} shows, the correction to the tree level 
formula, $\hat Z$, which is what is usually quoted as the Z-factor,
differs from unity by less than  than two per cent at $\kappa=0.123$.

We also attempted to measure the axial vector current renormalization
using the chiral Ward identity \cite{CHIRALWI}.
With 50 lattices, we only had a signal at $\kappa=0.118$, where
we found $Z_A=1.210(8)$.  This is again quite close to the pure
kinetic result of 1.237.  Dividing out the kinematic factors,
this would give $\hat Z_A=0.978$.

At $\beta=6.0$, Ref. \cite{MPSV} found $\hat Z_V=0.824$, $\hat Z_A=1.09$ for
the usual clover action, and improved operators.

\begin{table}
\begin{tabular}{|c|l|l|l|l|l|}
\hline
$\kappa$ & $Z^{cons}$  & $Z^{loc}$  &  $Z^{loc}/Z^{con}$  & 
${{1-6\kappa}\over{2\kappa}}$ \\
\hline
   0.118 &   1.020(1) & 1.253(2) & 1.228 & 1.237 \\
   0.120 &   1.032(2) & 1.185(4) & 1.148 & 1.166 \\
   0.122 &   1.054(5) & 1.129(6) & 1.071 & 1.098 \\
   0.123 &  1.071(6) & 1.127(8) & 1.052 & 1.065  \\
\hline
\end{tabular}
\caption{Vector current renormalization factors, $C_{sw}=1.2$, $\beta=5.7$.}
\label{tab:zv}
\end{table}

 It would be very interesting to measure the mixing
of left- and right-handed operators (for $B_K$) in this action.

\section{Conclusions}
We have shown that via a combination of fattening the links and
tuning the magnitude of the clover term,
 it is possible to optimize the chiral behavior of
the clover lattice fermion action.  The example we presented
reduces the spread in the low energy real eigenmodes by about a factor of three
in units of the squared pion mass, compared to the usual Wilson action.
We fixed the fattening and varied the clover coefficient, but it is clear
that a real optimization would involve varying both factors.

The same procedure can be applied to the usual clover action, without
fat links. Then,
 the value of the clover term $C_{sw}$ which minimizes the
spread of real eigenmodes due to instantons is so large, that  new
kinds of configurations become exceptional. Their
 singular behavior is unrelated to
fermion modes sitting on individual instantons. It is most likely
related to short distance vacuum fluctuations.
Thus this choice of action  does not represent a good choice for 
improvement.

The method of construction exposes an apparent fundamental upper limit
for the lattice spacing in a QCD simulation of about  0.2 fm. At larger
lattice spacing, the doubler modes and the low energy modes do not show
a clean separation, and the mechanism of chiral symmetry breaking 
is qualitatively different than in the continuum.
The Wilson and clover actions, and the
 one improved action we tested, the hypercubic action of Ref. \cite{HYPER},
failed at  lattice spacing  0.24 fm.
This test does not apply to actions involving heavy quarks, for which chiral
symmetry is (presumably) not important, but before any other improved action
can be said to reproduce continuum light hadron
physics at some lattice spacing,
it should show a separation between the spectrum of
its near low energy modes and its doubler modes.

The ``fat clover'' action we have presented in this paper, as an example
 of an action with improved chiral properties, has many other nice 
properties as well.  It appears to exhibit scaling of hyperfine
splittings at 0.2 fm. The action is quite insensitive to the UV
behavior of the underlying gauge field.
It has Z-factors for simple matrix elements which
are very close to unity. 
And finally, the amount of resources required to construct propagators
is halved compared to the usual clover action, at equivalent parameter
values.  It would be trivial to modify any existing clover code to
improve it in the way we have described.  Of course, its kinetic
properties, including the dispersion relation,  artifacts
at large $am_q$, and power law scaling violations in matrix elements,
are unchanged from the standard clover action's.

\section*{Acknowledgements}
We are greatly indebted to Hank Thacker who provided us
the exceptional configurations reported in Ref. \cite{FNALINST}.
We would like to thank the UCLA TEP group for granting us part
of the computer time used for this work.
Part of the computing was done on the 
Origin 2000 at the University of California, Santa Barbara.
This work was supported by the U.S. Department of 
Energy.

\newcommand{\PL}[3]{{Phys. Lett.} {\bf #1} {(19#2)} #3}
\newcommand{\PR}[3]{{Phys. Rev.} {\bf #1} {(19#2)}  #3}
\newcommand{\NP}[3]{{Nucl. Phys.} {\bf #1} {(19#2)} #3}
\newcommand{\PRL}[3]{{Phys. Rev. Lett.} {\bf #1} {(19#2)} #3}
\newcommand{\PREPC}[3]{{Phys. Rep.} {\bf #1} {(19#2)}  #3}
\newcommand{\ZPHYS}[3]{{Z. Phys.} {\bf #1} {(19#2)} #3}
\newcommand{\ANN}[3]{{Ann. Phys. (N.Y.)} {\bf #1} {(19#2)} #3}
\newcommand{\HELV}[3]{{Helv. Phys. Acta} {\bf #1} {(19#2)} #3}
\newcommand{\NC}[3]{{Nuovo Cim.} {\bf #1} {(19#2)} #3}
\newcommand{\CMP}[3]{{Comm. Math. Phys.} {\bf #1} {(19#2)} #3}
\newcommand{\REVMP}[3]{{Rev. Mod. Phys.} {\bf #1} {(19#2)} #3}
\newcommand{\ADD}[3]{{\hspace{.1truecm}}{\bf #1} {(19#2)} #3}
\newcommand{\PA}[3] {{Physica} {\bf #1} {(19#2)} #3}
\newcommand{\JE}[3] {{JETP} {\bf #1} {(19#2)} #3}
\newcommand{\FS}[3] {{Nucl. Phys.} {\bf #1}{[FS#2]} {(19#2)} #3}


\begin{thebibliography}{9}

\bibitem{VACUUM}
Cf. the review talk of P. van Baal at Lattice 97, Nucl. Phys, 
B (Proc. Suppl.) 63 (1998) 126.

\bibitem{NEGELE}
J.~Negele, plenary talk at Lattice 98, hep-lat/9810053.

\bibitem{SU3INST}
A.~Hasenfratz and C.~Nieter, hep-lat/9806026.

\bibitem{SU3INST_alt}
M.~Chu, J.M.~Grandy, S.~Huang, and J.W.~Negele, \PR{D49}{94}{6039}, 
D.A.~Smith, and M.J.~Teper,\PR{D58}{98}{014505},  hep-lat/9801008, 
P.~de~Forcrand, M.~Garcia~Perez, J.E.~Hetrick, and I.O.~Stamatescu,
contributed to the 31st International Ahrenshoop Symposium on hte 
Theory of Elementary Particles, Buckow, Germany, 2-6 Sep 1997;
hep-lat/9802017.


\bibitem{IILM}
Cf. D.~Diakanov, Lectures at the Enrico Fermi School in Physics,
Varenna, 1995, hep-ph/9602375;
T.~Sch\"afer and E.~V.~Shuryak, \REVMP{70}{98}{323}.

\bibitem{FNALINST}
W.~Bardeen, A.~Duncan, E.~Eichten, G.~Hockney and H.~Thacker,
\PR{D57}{98}{1633};
W.~Bardeen, A.~Duncan, E.~Eichten and H.~Thacker,
\PR{D57}{98}{3890};
hep-lat/9606002.


\bibitem{HNL}
P.~Hasenfratz, V.~Laliena and F.~Niedermayer,\PL{B427}{98}{125},
 hep-lat/9801021;
P.~Hasenfratz, \NP{B525}{98}{401}, hep-lat/9802007.


\bibitem{DESY96173}
M.~G\"ockeler, et al. \PL{B391}{97}{388}.



\bibitem{APEBlock}
 M.~Falcioni, M.~Paciello, G.~Parisi, B.~Taglienti,
\NP{B251[FS13]}{85}{624}.
M. Albanese, et.~al. \PL{B192}{87}{163}.





\bibitem{MITCOOL}
M.-C.~Chu, J.~M.~Grandy, S.~Huang and J.~W.~Negele, \PR{D49}{94}{6039}.

\bibitem{MILC}
T.~Blum, et al., \PR{D55}{97}{1133}.

\bibitem{ORT}
K. Orginos and D.~Toussaint, hep-lat/9801020.

\bibitem{SINCLAIR}
J.-F. Laga\"e and D.~K. Sinclair, Nucl. Phys. B(Proc. Suppl.) 63 
(1998) 892.; hep-lat/9806014.


\bibitem{COLOINST}
T.~DeGrand, A.~Hasenfratz, T.G.~Kov\'acs,
\PL{B420}{98}{97}.



\bibitem{ALLFP}
W. Bietenholz and U.~J. Wiese, \NP{B464}{96}{319};
 T. DeGrand, A. Hasenfratz, P. Hasenfratz, P. Kunszt, F. Niedermayer, 
Nucl. Phys. B (Proc. Suppl.) 53, 1997, 942;
W.~Bietenholz, et al., Nucl. Phys. B(Proc. Suppl.) 53 (1997) 921;
K.~Orginos, et al.,  hep-lat/9709100,
Nucl. Phys. B (Proc. Suppl.) 63 (1998) 904;
C.~B. Lang and T.~K. Pany, hep-lat/9707024,
\NP{B513}{98}{645}.



\bibitem{HYPER}
T.~DeGrand, \PR{D58}{98}{094503}.


\bibitem{SMITVINK}
J.~Smit and J.~Vink, \NP{B286}{87}{485}, J.~Vink, \NP{B307}{88}{549}.

\bibitem{SIMMA} H.~Simma, D.~Smith, Low-lying eigenvalues of the improved 
Wilson-Dirac operator in QCD, hep-lat/9801025.

\bibitem{SCRI_smooth} R.~G.~Edwards, U.~M.~Heller, R.~Narayanan,
\NP{B522}{98}{285}.


\bibitem{SCRI_spectralflow}R.~G.~Edwards, U.~M.~Heller, R.~Narayanan,
\NP{B535}{98}{403}, hep-lat/9802016.



\bibitem{STRONG}
H.~Kluberg-Stern, A.~Morel, and B.~Petersson, \PL{B114}{82}{152}.

\bibitem{LUSCHER} M.~L\"uscher, S.~Sint, R.~Sommer, 
and P.~Weisz, \NP{B502}{96}{365}

\bibitem{GW}
P.~Ginsparg and K.~Wilson, \PR{D25}{82}{2649}.

\bibitem{HEMCGC}
K.~M. Bitar, et al., \PR{D42}{90}{3794}.

\bibitem{DOUGFPI}
C. Bernard, et al., 
\PR{D45}{92}{3854}.

\bibitem{BUTLER}
F.~Butler, H.~Chen, J.~Sexton, A.~Vaccarino, and D.~Weingarten,
\NP{B430}{94}{179}.




\bibitem{APE63} M. Guagnelli, et al., \NP{B378}{92}{616}.

\bibitem{ALPHA}
 M.~G\"ockeler, et.~al., hep-lat/9707021, \PR{D57}{98}{5562}.

\bibitem{SCRI}
R.~G.~Edwards, U.~M.~Heller and T.~R. Klassen, \PRL{80}{98}{3448},
 hep-lat/9711052.

\bibitem{ZFACTOR}
C.~Bernard, J.~Labrenz, and A.~Soni, \PR{D49}{94}{2536};
G.~P.~Lepage and P.~Mackenzie, \PR{D48}{93}{2250}.


\bibitem{CHIRALWI}
M.~Bochicchio, et. al., \NP{B262}{85}{331};
L.~Karsten and J.~Smit, \NP{B183}{81}{103}.
L.~Maiani and G.~Martinelli, \PL{B178}{86}{285}.

\bibitem{MPSV}
G.~Martinelli, S.~Petrarca, C.~T.~Sachrajda, \PL{B311}{93}{241}.

\end{thebibliography}
\end{document}